%% file: main.tex
\documentclass{article}

\usepackage{arxiv}
\usepackage{times}
\usepackage{natbib}

\input{math_commands.tex}

\usepackage{hyperref}
\usepackage{url}
\usepackage{graphicx}
\graphicspath{{fig/}}
\usepackage{amsmath,amssymb}
\usepackage{amsthm}
\newtheorem{proposition}{Proposition}
\newtheorem{definition}{Definition}
\newtheorem{corollary}{Corollary}
\usepackage{multirow}
\usepackage{booktabs}
\usepackage{algorithm}
\usepackage{algpseudocode}
\usepackage{xcolor}      % required by \SS in commands.tex
\usepackage{subcaption}

\input{commands.tex}

\title{Guiding Where the Antibody Binds:\\Amortised Cross-Subsample Particle Steering for
Antibody--Antigen Co-Folding}

\renewcommand{\shorttitle}{CAST: Amortised Cross-Subsample Particle Steering}

\author{%
  Luwei Wang$^{1}$ \quad Ferran Gonzalez$^{2}$ \quad Xinheng He$^{2}$ \quad
  Raimon Fabregat$^{2}$ \quad Leon Gerard$^{2}$ \\[0.5em]
  \normalfont\normalsize
  $^{1}$School of Informatics, University of Edinburgh, UK \qquad $^{2}$AstraZeneca
}
\date{}

\begin{document}
\maketitle

\begin{abstract}
Co-folding models have advanced antibody-antigen structure prediction, but a dominant failure mode persists. They often dock the antibody onto the wrong epitope. Docking has two problems. A sampler must produce an acceptable complex at the correct interface, and a verifier must then rank it above the alternatives. We address the first, \emph{site coverage}, since a verifier cannot select an epitope the sampler never proposed. Drawing more seeds or samples adds variety around the pose the model already favours, but does not relocate it.
We introduce \textbf{CAST} (\emph{Cross-subsample Amortised Site Tilting}), which steers diffusion trajectories toward a binding \emph{site} using a reward built from a candidate epitope map, training-free and with no reference structure. The sampler approximates an ideal target whose tilt provably changes the site marginal but not the pose at a site. Amortising $K$ alignment-subsample latents into one steered pool cuts the cost of exploring them from $\mathcal{O}(KN)$ to $\mathcal{O}(N+K)$ for a pool of $N$ particles.
At a matched $720$ structures and $4$ conditionings, CAST holds an acceptable pose on $36$ of the $50$ difficulty-balanced CFA50 targets against $26$ for Best-of-$N$, and raises mean max DockQ from $0.322$ to $0.373$, a gain of $15.8\%$. On $21$ targets that a $1{,}000$-seed run fails to reach CAPRI-medium quality, CAST reaches mean max DockQ $0.370$ at $720$ structures and brings $18$ of the $21$ to an acceptable pose against $9$ at the same budget. Best-of-$N$ flattens, spending $5.3\times$ the compute and trailing by $20.1\%$. Raising the latent count alongside the samples carries this to $0.457$, beyond every Best-of-$N$ budget. On CFA50 CAST also matches the best Best-of-$N$ allocation of $720$ structures, at $1.5\times$ less compute. Such steering is a reallocation, changing little where the base model already succeeds and much where it does not.
\end{abstract}

\section{Introduction}
\label{sec:intro}

A predicted antibody--antigen complex is useful only if the antibody is docked at the right epitope,
the patch a design campaign builds against \citep{germinal2025} and an affinity model reads
\citep{abrank2025}, and the system can recognise it. Co-folding therefore has two bottlenecks.
The sampler must produce an acceptable complex at the correct interface, and a verifier must then
rank it above the incorrect alternatives.

This work addresses the first, which we call \emph{site coverage}. A verifier can reorder structures
already present in a pool. It cannot, however, select a correct epitope that the sampler never
proposed. The two problems are also separate in practice, since the interface confidence scores used
to rank such structures are themselves unreliable \citep{dunbrack2025ipsae}. We therefore isolate sampling coverage and
report what the pool makes available, the number of targets holding an acceptable complex and the
best structure present, $\mathrm{DockQ}^{\max}_{\mathrm{abag}}$
\citep{mirabello2024dockq2}, not what a ranker returns.

Coverage is where co-folding currently fails. Models predict antibody--antigen complexes directly
from sequence \citep{abramson2024af3,wohlwend2025boltz1,protenix2025,candido2026esm,protenixv2}, yet
AlphaFold3 fails to dock $65\%$ of them at one seed \citep{hitawala2025af3ab}, and accuracy drops further
without a close structural analogue in training \citep{fromm2026abag}. Across $1{,}628$ complexes,
methods place the antibody incorrectly and converge on the \emph{same redundant binding mode}
\citep{clifford2026pocket}. That wrong site is not a near miss. A structure on a different patch
shares almost no native contact, so its DockQ is near zero. No local move reaches a distant patch and
repairs it, and the model will not offer an alternative. 
This calls for two capabilities, naming a site and steering the sampler to it.

\textit{Spending more supplies neither.} \citet{fromm2026abag} find accuracy rises roughly linearly in
the \emph{logarithm} of the sample count, because independent draws deepen the mode the model favours
instead of relocating it. Of the $52$ targets we collected, $18$ still hold nothing acceptable at
$1{,}000$ unsteered structures each, and $7$ recover no native contact at all (\S\ref{app:failmode}).

\textbf{Related work.} \textit{Conditioning names a site, but pays for each one.} A Chai-1 \citep{chai1} pocket restraint
from a B-cell epitope predictor improves accuracy and diversity \citep{clifford2026pocket}, so the
modes are reachable inside a frozen model. But the restraint enters before the trunk and is fixed for
the trajectory, so $S$ candidate sites cost $S$ trunk passes. ACER \citep{acer2026} pays it
twice, once per decoy configuration it occludes the dominant pocket with, and again per candidate
pocket when it turns each one into a restraint to rank under.
Inside the diffusion loop a site costs no trunk pass of its own. \textit{Steering works there, but names no site.} It diversifies the \emph{conformation} a complex
adopts, not the \emph{site} where it binds, by weakening the alignment signal
\citep{afsample2,afsample3}, rescaling trunk latents \citep{boltzsample2026,confornets}, or guiding
and repelling in the sampler \citep{conformix2025,corso2024particle,conforflux2026}. Those
that repel score by deviation from a reference, so the mass they move lands wherever
the base model ranks next, never on a site we name. ACER \citep{acer2026} repels in site space
instead, off pockets already visited, which makes that explicit, since each round returns whichever
pocket the model prefers next rather than one we choose. The few that aim somewhere use a target correct
by construction that we lack, an experimental measurement \citep{embedopt} or physical prior
knowledge \citep{forceguided}.
Twisted sequential Monte Carlo \citep{wu2023practical} and Feynman--Kac steering
\citep{singhal2025fk} are training-free and backbone-agnostic but prescribe \emph{how} to steer, not
\emph{what} to steer toward. We are not aware of a method that carries a binding site as an
\emph{attractive} objective inside the diffusion loop (\S\ref{sec:related}).

We introduce \textbf{CAST}, Cross-subsample Amortised Site Tilting, which supplies that objective. It
is training-free, builds on the twisted sampler, and steers it toward a binding site from inside the
loop. Without a
reference structure the only site signal is a predicted epitope, itself an
open problem \citep{liu2024asep}. CAST therefore treats its map as a direction, not an answer, and
sweeps a grid of candidates. A site is a destination carried by a particle, and so is the
conditioning latent, so site hypotheses and the $K$ MSA subsamples share one pool instead of $K$
pools. At the same $720$ structures this brings $36$ of $50$ targets to an acceptable pose against $26$. 

\paragraph{Contributions.} \textbf{(i) A site collective variable and its tilt}
(\S\ref{sec:cast}), a reward on the paratope's position alone, hinged at an arrival radius and swept
over \emph{umbrellas}, one per target separation~radius~$d^\star$. It is a soft prior, so CAST reaches epitopes a displaced map candidate would exclude.
\textbf{(ii) Cross-subsample amortisation} (\S\ref{sec:cast}). We make the conditioning latent part
of the particle state, so the $K$ subsamples share one pool. That cuts the cost from
$\mathcal{O}(KN)$ to $\mathcal{O}(N+K)$, and stratification keeps the candidate grid occupied at any
$K$. \textbf{(iii) A reference-free epitope map} (\S\ref{sec:map}) from pool binding modes plus a
farthest-point surface tiling, without ground truth. \textbf{(iv) A pose-invariance guarantee}
(\S\ref{sec:theory}). Under a site-measurable reward the ideal target preserves the pose conditional at a fixed
site, at every temperature. The tilt thus controls the site marginal alone, and leaves
the pose at that site to the base model. \textbf{(v) Empirical gains and the axis they
scale on} (\S\ref{sec:results}). At a matched $720$ structures the gain sits entirely on the targets
Best-of-$N$ does not reliably solve, where CAST brings $16$ of $29$ to threshold against $6$. It
reaches threshold on $36$ targets against $30$, $27$, $24$ and $22$ for four steered baselines. The gain
scales as latents and samples rise together, $0.370$ to $0.457$, while Best-of-$N$ flattens at every budget. 
CAST also transfers to a second backbone without retuning (\S\ref{app:opendde}).

\section{Method}
\label{sec:method}

\subsection{Preliminaries}
\label{sec:prelim}

A co-folding model maps sequences to atomic coordinates in two stages
\citep{abramson2024af3,protenixv2}. A \emph{trunk} embeds the input, including a multiple-sequence
alignment, and produces a conditioning representation $z$. A diffusion \emph{decoder} then samples
$x\sim p_\theta(x\g z)$ by integrating a reverse process from noise. The decoder is variance-exploding \citep{karras2022edm}, with $x_t=x_0+\sigma_t\varepsilon$ for
$\varepsilon\sim\normal(0,\I)$. Writing $p_t(x_t\g z)$ for the marginal at noise level $\sigma_t$ and
$D_\theta$ for the denoiser, Tweedie's identity relates the denoiser to the score
\citep{vincent2011denoising,karras2022edm},
\begin{equation}
\hat x_0\;=\;D_\theta(x_t,\sigma_t\g z)\;\approx\;\E[x_0\g x_t,z]
\;=\;x_t+\sigma_t^2\,\nabla\log p_t(x_t\g z),
\label{eq:tweedie}
\end{equation}
with equality in the first step for a Bayes-optimal denoiser. The sampler is the stochastic update
of \citet{karras2022edm}. A step recenters and randomly rotates the coordinates
\citep{abramson2024af3}, injects noise to raise the level from $\sigma_t$ to $\hat\sigma_t\ge\sigma_t$,
\begin{equation}
\tilde x_t=x_t+\beta_t\varepsilon_t,\qquad
\beta_t=\sqrt{\hat\sigma_t^2-\sigma_t^2},\qquad \varepsilon_t\sim\normal(0,\I),
\label{eq:churn}
\end{equation}
denoises once at that raised level, then descends deterministically to
$x_{t-1}=\tilde x_t+(\sigma_{t-1}-\hat\sigma_t)(\tilde x_t-\hat x_0)/\hat\sigma_t$, along
$-\hat\sigma_t\nabla\log p_{\hat\sigma_t}(\tilde x_t\g z)$ by \eqref{eq:tweedie}. The rigid-frame
augmentation is an isometry, so after it the injection in \eqref{eq:churn} is the only stochastic
perturbation of the diffusion state.

\textbf{Reward tilting.} Biasing such a decoder toward high-reward structures is usually posed as
sampling from a reward-tilted target, $q(x)\propto p_\theta(x\g z)\exp(\lambda r(x))$ for a reward $r$
and a strength $\lambda>0$. Reaching it needs the \emph{time marginals} of $q$, which carry an
intermediate potential $h_t(x_t)=\E[\exp(\lambda r(x_0))\g x_t,z]$ under the endpoint posterior,
since $q_t\propto p_t\,h_t$ makes the tilted score the base score plus $\nabla\log h_t$. That is
exact and intractable, and a structural reward is in any case defined on clean coordinates not on 
a noisy iterate. Evaluating it on the denoised estimate answers both at once
\citep{chung2023dps,wu2023practical}, giving the plug-in potential
$\hat h_t=\exp(\lambda r(\hat x_0))$, whose gradient carries the denoiser Jacobian
$J_{D_\theta}(x_t)=\partial\hat x_0/\partial x_t$,
\begin{equation}
\nabla_{x_t}\log\hat h_t=\lambda\,J_{D_\theta}(x_t)^{\!\top}\,\nabla_{\hat x_0} r(\hat x_0).
\label{eq:plugin}
\end{equation}
Twisted sequential Monte Carlo treats such a tilt as a \emph{proposal}, not as the answer and
corrects it by reweighting a population of particles \citep{wu2023practical,singhal2025fk}. Reserving
$q$ for the target, a guided sampler draws the injection of \eqref{eq:churn} from a shifted mean
$x_t+g$ instead of $x_t$ and changes nothing else. Writing $r_t=r(\hat x_0)$ for the reward read on
the denoised estimate at step $t$, both factors of the weight are then closed-form,
\begin{equation}
w_t\;\propto\;\exp\big(\!-[\norm{g}^2+2\beta_t\langle\varepsilon_t,g\rangle]/2\beta_t^2\big)\;\cdot\;\exp\big(\lambda(r_{t-1}-r_t)\big),
\label{eq:ratios}
\end{equation}
an exact likelihood ratio for the Gaussian \emph{injection kernel}, paying for the shifted
injection drawn in place of the decoder's own, times a \emph{potential ratio} imposing the reward
tilt. Reverse sampling runs from $t$ to $t-1$, so the potential ratio exceeds one exactly when the
reward improves. Particles are resampled in proportion to these weights when their effective
sample size falls too low, which turns weight into pool composition.

\subsection{CAST}
\label{sec:cast}

\begin{figure}[!ht]
\begin{center}
\includegraphics[width=\textwidth]{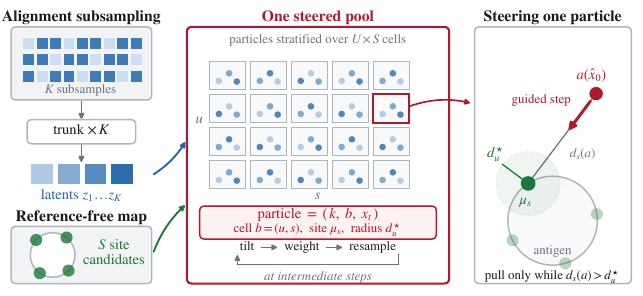}
\end{center}
\caption{\textbf{CAST in one pool.} $K$ alignment subsamples and the reference-free map of
\S\ref{sec:map} feed one steered pool, whose particles are stratified over the $U\times S$ cells.
\textbf{Right}, one cell opened up. The guided step displaces the paratope centroid toward the
candidate centroid $\mu_s$, and vanishes once it is within the arrival radius $d^\star_u$.}
\label{fig:method}
\end{figure}

CAST augments the particle state with a destination. Each particle carries the candidate site it
targets and the conditioning it was drawn from, both fixed for the whole trajectory. One pool
therefore covers the full candidate grid across all $K$ MSA subsamples. A particle is the triple
$(k,\;b,\;x_t)$, with diffusion state $x_t$, an index $k$ into the conditionings $z_1,\dots,z_K$,
and cell $b=(u,s)$ pairing a candidate site $s$ with an arrival radius $u$, or \emph{umbrella}.
Guidance displaces each particle toward its cell, and resampling selects among them
(Figure~\ref{fig:method}). Resampling may reallocate $k$, whereas stratification fixes the cells.
Both act on the antibody's position, not its pose.

\textbf{Site and pose.} The \emph{paratope} is fixed before sampling as the IMGT CDR residues under
ANARCI numbering \citep{dunbar2016anarci}, so it is defined on any partly denoised structure. We
parameterise a complex by a site coordinate $a$, an orientation $R$ and internal coordinates $c$. The
site $a$ is the paratope centroid in the antigen frame, so a global rigid motion of the complex leaves
it fixed, and $c$ is expressed about that centroid, so a change of internal geometry cannot move it.
The chain rule then gives each latent its own factorisation
\begin{equation}
x=(a,R,c),\qquad
p(a,R,c\g z_k)=p(a\g z_k)\,p(R,c\g a,z_k).
\label{eq:factor}
\end{equation}
The first factor is \emph{where} the antibody sits, the second is \emph{how} it is arranged once
placed. Proposition~\ref{prop:invariance} shows that a reward depending on $x$ only through $a$
leaves the second factor invariant.

\textbf{Joint target over structures and conditionings.} Different alignment subsamples produce
different conditioning representations \citep{afsample2,afsample3}, so $z_k$ indexes a genuinely
\emph{different} conditional and not another draw from a single one, so the choice of conditioning becomes part of the
sampling problem instead of a setting fixed beforehand. Given candidate centroids $\mu_s$ from the
map of \S\ref{sec:map}, arrival radii $d^\star_u$, and a uniform prior $\rho_k=1/K$ over latents, each
cell $b$ defines a reward-tilted target on the augmented state
\citep{wu2023practical,singhal2025fk},
\begin{equation}
q_b(k,x)\;\propto\;\rho_k\,p_\theta(x\g z_k)\exp\!\big(\lambda\,r_b(a)\big),
\qquad
r_b(a)=-\,\ell_\delta\!\big(\max(0,\;d_s(a)-d^\star_u)\big),
\label{eq:tilt}
\end{equation}
where $\ell_\delta$ is a Huber function and $d_s(a)=\norm{a-\mu_s}$ is the paratope-to-candidate
separation. We write $r_b(x)=r_b(a(x))$ for the reward read on a structure, so $r_b$ sees the site
coordinate and nothing else, the property Definition~\ref{def:site} names. The hinge makes $d^\star_u$ an arrival radius, not a target
point, so a particle already inside it feels neither reward nor gradient, and it is also what keeps
the strong-tilt limit a conditioning instead of a collapse (Corollary~\ref{cor:ball}). We sweep $u$
over several paratope-to-site distances (\S\ref{app:umbrella}). Marginalising the latent tilts the \emph{mixture} of the $K$ conditionals,
\begin{equation}
q_b(x)\;\propto\;p_{\mathrm{mix}}(x)\,\exp\!\big(\lambda\,r_b(a)\big),
\qquad
p_{\mathrm{mix}}(x)=\textstyle\sum_{k=1}^{K}\rho_k\,p_\theta(x\g z_k),
\label{eq:mixture}
\end{equation}
so a single particle system can reallocate decoder trajectories across conditionings. $K$ separate
pools do not perform this reallocation, because their allocations are fixed in advance, and matching
\eqref{eq:mixture} would additionally require weighting the pools by their per-latent normalising
constants. Particles are partitioned over the cell grid
$\mathcal{U}\times\mathcal{S}$, indexing $U$ arrival radii against $S$ candidates, each cell holding a
fixed share of the pool and targeting its own $q_b$. Reserving budget cell by cell is what lets the
pool reach sites the base model would not visit on its own.

\textbf{One pool over $K$ latents.} Both designs evaluate the trunk $K$ times.
Steering separately runs $K$ pools of $N$ particles each, whereas CAST defers the decoder until all
$K$ latents are in hand and runs one pool of $N$ that every latent shares. Separate steering then
needs $KN$ decoder trajectories where CAST needs $N$, so a latent is an addition for CAST,
$\mathcal{O}(N+K)$, and a multiplication for a separate-model arm, $\mathcal{O}(KN)$.
The multiplication is stratification's, since each separate pool must occupy the whole cell grid
on its own. A cell must also hold enough particles to resample with its occupancy divided among the
latents, so $N$ and $K$ are floored and capped together (\S\ref{app:assign}, with structure counts
in \S\ref{app:compute}).

\textbf{Guided proposal.} CAST steers by displacing the mean of the Gaussian injection
in \eqref{eq:churn} for particle $i$ by
$g^i_t=\mathrm{clip}\big[\eta_t\,\nabla_{\hat x_0} r_b(\hat x_0^i)\big]$, under a step $\eta_t$
annealed with the noise level and a size-invariant cap (\S\ref{app:guidance}). Because the churn
precedes this step's denoise, it is read on the preceding step's denoised estimate and adds no
decoder call of its own. Each scoring event below costs
four denoiser passes per particle (\S\ref{app:compute}). Algorithm~\ref{alg:cast} in
\S\ref{app:algorithm} states the whole sampler.
The reward reads a centroid separation, so $\nabla_{\hat x_0}r_b$ is available in closed form, and
we take the Jacobian-free straight-through approximation $J_{D_\theta}\approx\I$ in
\eqref{eq:plugin} \citep{bengio2013ste,moufad2026vjpfree}, so the guided step needs no backward
pass through the decoder or the reward. The injection is the step's only stochastic perturbation and
the shift is applied only there, so \eqref{eq:ratios} is exact at $g=g^i_t$.

\textbf{Weights and resampling.} Weights accumulate both factors of \eqref{eq:ratios}, the potential ratio only at a sparse set of
\emph{scoring events} $m=0,\dots,M$, with $r^i_m$ the reward of particle $i$ at event $m$. The
potential enters through increments, and these telescope to
$e^{\lambda(r^i_M-r^i_0)}$. We add the initial term back once, giving $e^{\lambda r^i_M}$, which keeps
the baseline differences between latents and needs $K>1$ (\S\ref{app:amortisation}). Resampling runs \emph{within} cells and only inside an early window $\mathcal{W}$, and the two
restrictions protect different diversity. Within cells protects \emph{site} diversity, since a global
resample would let one cell take another's budget, at the price of capping each candidate at a $1/S$
share of the pool. Early protects \emph{pose} diversity, since once the site has settled a resample
can only clone, and under the ideal target a site-only reward prefers no pose over another. Weights
are consumed only inside $\mathcal{W}$ and the ensemble is returned unweighted, while guidance
continues past $\mathcal{W}$ as an unweighted drift, so the asymptotic guarantee of
\citet{wu2023practical} does not apply.

\subsection{A site tilt preserves the pose conditional}
\label{sec:theory}

We now make the separation between site and pose precise. Definition~\ref{def:site} names the class
of rewards it requires, Proposition~\ref{prop:invariance} states the invariance, and
Corollary~\ref{cor:ball} gives the strong-tilt limit.

\begin{definition}[site-measurable reward]
\label{def:site}
A reward is \emph{site-measurable} if it depends on a structure $x=(a,R,c)$ only through its site
coordinate, $r(x)=r(a)$.
\end{definition}

The reward of \eqref{eq:tilt} is site-measurable. A reward reading interface contacts,
clashes, orientation or internal geometry is not site-measurable, and none of what follows applies
to those.

\begin{proposition}
\label{prop:invariance}
Let $q$ be the tilt of \eqref{eq:tilt} with any site-measurable reward. For every $\lambda$ the pose
conditional is unchanged both within each latent and, after marginalising $k$, for the mixture,
\begin{equation*}
q(R,c\g a,z_k)=p(R,c\g a,z_k)\ \ \text{for every }k,
\qquad\text{and}\qquad
q(R,c\g a)=p(R,c\g a).
\end{equation*}
\end{proposition}
\begin{corollary}[the $\lambda\to\infty$ limit is a conditioning, not a collapse]
\label{cor:ball}
Write $\mathcal{B}_b=\{a\in\reals^3:\norm{a-\mu_s}\le d^\star_u\}$ for the arrival ball, and assume
$p_{\mathrm{mix}}(a\in\mathcal{B}_b)>0$. With
the hinge of \eqref{eq:tilt} the tilt factor tends to an indicator on $\mathcal{B}_b$, so the
mixture target of \eqref{eq:mixture} tends to a conditioning of the same mixture,
%\begin{equation*}
$q_b(x)\;\xrightarrow[\lambda\to\infty]{}\;p_{\mathrm{mix}}\big(x\g a\in\mathcal{B}_b\big).$
%\end{equation*}
The same holds at a fixed latent whenever $p_\theta(a\in\mathcal{B}_b\g z_k)>0$, with
$q_b(x\g k)\propto p_\theta(x\g z_k)e^{\lambda r_b(a)}$ tending to
$p_\theta(x\g a\in\mathcal{B}_b,z_k)$.
\end{corollary}

Both proofs are in \S\ref{app:proofs}. The hinge creates the zero-reward region
inside the ball, and without it a larger $\lambda$ would favour ever smaller separations
instead of conditioning on an arrival region. The finite particle system approximates $q_b$ through a plug-in potential, a
straight-through gradient, and a truncated window. Only the last costs exactness, and
\S\ref{app:algorithm} states the cost of each.

\subsection{Reference-free candidate map}
\label{sec:map}

The map is built without ground truth from two ingredients with complementary coverage. Pool
\emph{modes} are the binding-site modes of an unsteered pilot pool, antibody-conditioned by
construction and requiring no reference structure. \emph{Tiles} come from a farthest-point covering of the
antigen surface, so every residue lies within the covering radius of some seed. The map lists the
leading modes, then tiles ranked by their overlap with those modes, up to a fixed candidate budget.
\S\ref{app:map} gives the construction, and sizes are in \S\ref{sec:experiments}.

\section{Experimental setup}
\label{sec:experiments}

\textbf{Evaluation set.} 
\textbf{CFA50} comprises $50$ antibody- and nanobody--antigen complexes from the CoFold Arena panel
\citep{cofoldarena}, which admits only structures released after 9 September 2024 and so gives a
clean time split. Two different quantities are at work. CoFold Arena scores a target by
$\mathrm{DockQ}_{\mathrm{abag}}$ on the model's single \emph{best-ranked} structure. We screened
$80$ panel entries, stratifying on format, size and antigen identity (\S\ref{app:datasets}). We
report the pool \emph{maximum} instead, so we then ran an unsteered pool of $200$ structures per
candidate and kept the $50$ that
\emph{balance} whether any structure reaches $\mathrm{DockQ}_{\mathrm{abag}}\ge0.23$.
\textbf{Collected 21-hard} is a second set of $21$ complexes, used to see how performance
scales with budget on hard cases. We curated it from PDB depositions released after 30 June 2025 and novel against
earlier entries, under the structure and interface filters of \S\ref{app:datasets}. CDR-sequence
clustering then keeps only a cluster representative, so they are not near-duplicates of one
another. From these we kept the targets a $100$-seed run of the base model leaves unsolved, then the
$21$ that a $1000$-seed run still leaves below the CAPRI-medium threshold $0.49$
\citep{mendez2003capri}. Selection and evaluation use different runs. The $1000$-seed run selects the targets and a
separate $50$-conditioning pool supplies every Best-of-$N$ number, so the draws scored are
disjoint from those that chose the set.

\textbf{Baselines.} Every baseline gets CAST's budget and conditioning count, $720$ structures as $4$
MSA subsamples $\times$ $180$ samples. \emph{Best-of-$N$} (BoN) is the unsteered control and the reference
for all comparisons. Since a matched budget can still be spent badly, we also sweep its allocation
(\S\ref{app:alloc}). Four further baselines each steer by a different
mechanism. \emph{Feynman-Kac steering} (FK) \citep{singhal2025fk} applies the same
reward and resampling without the gradient, at a single $d^\star=12$\,\AA{}. \emph{Noise search}
(NS) \citep{jajal2025evo} runs an evolutionary search over the initial noise. We run their CoSyNE at
a population of $12$ over $7$ generations, so that it spends the same $180$ evaluations per conditioning. \emph{Boltz-Perturb} (BP) \citep{boltzperturb2025} is conditioning perturbation with no
objective and no resampling. \emph{AF3-ReD} (ReD) \citep{af3red2025} repels each conditioning from
the structures already produced, over whole-protein C$\alpha$, also with no objective and no
resampling.

\textbf{Implementation.} All arms run on Protenix-v2 \citep{protenixv2}, one backbone and one
checkpoint, $10$ recycles and $200$ diffusion steps, with MSAs subsampled to depth $1000$
independently per conditioning. Steering uses the reference-free hybrid map at a $12$-tile budget,
giving at most $S=15$ candidates, crossed with $U=4$ umbrella
radii $9,11,13,15$\,\AA{}, so the $\mathcal{U}\times\mathcal{S}$ grid holds at most $60$ resampling
cells. The map's modes are read from a $150$-structure unsteered pilot pool.
Runs use A100 GPUs and ground truth is used for scoring only.

\textbf{Metrics.} $\mathrm{DockQ}^{\max}_{\mathrm{abag}}$, the pool maximum of DockQ restricted to
the antibody--antigen interface \citep{basu2016dockq,mirabello2024dockq2}, is the primary measure and states what a pool makes
available, since ranking is a separate failure and existing confidence estimates often cannot identify the best
structure a pool contains \citep{fromm2026abag}. A
target \emph{reaches} $0.23$ when its $\mathrm{DockQ}^{\max}_{\mathrm{abag}}$ clears the
CAPRI-acceptable threshold \citep{mendez2003capri}, and the tables count targets that do, which
measures how many targets hold a usable structure at all. Compute is reported as PFLOPs per target, counting one trunk pass per
conditioning plus the denoiser and confidence passes each structure costs
(\S\ref{app:compute}), so two arms at the same structure count can differ in compute.

\textbf{Reporting protocol.} Every $\mathrm{DockQ}^{\max}_{\mathrm{abag}}$ we compare is at a
matched $720$ structures, and a BoN rung uses the $K$ lowest conditioning indices with
$720/K$ samples from each. At $K=4$ a rung holds every sample of its $4$ conditionings, so the value
is exact. Above $K=4$ a rung takes a subset of each conditioning's samples, and we report the exact
expectation of the maximum over those subsets (\S\ref{app:expect}). The scaling comparison
(\S\ref{sec:res-main}, Figure~\ref{fig:main4}(c,d)) states each budget, and
averages over every conditioning subset of that size instead of fixing the lowest indices. CFA50
targets are grouped by the fraction of $4$-conditioning draws in which BoN alone reaches
$\mathrm{DockQ}^{\max}_{\mathrm{abag}}\ge0.23$, computed exactly over every subset.
\emph{Reliably fails} at $\le10\%$, \emph{reliably solves} at $\ge90\%$, and \emph{borderline}
between (\S\ref{app:group}). Differences between arms are reported as means over targets, with
paired win counts.

\section{Results}
\label{sec:results}

\subsection{CAST against BoN, at a matched budget and under scaling}
\label{sec:res-main}

\begin{figure}[!ht]
\begin{center}
\includegraphics[width=\textwidth]{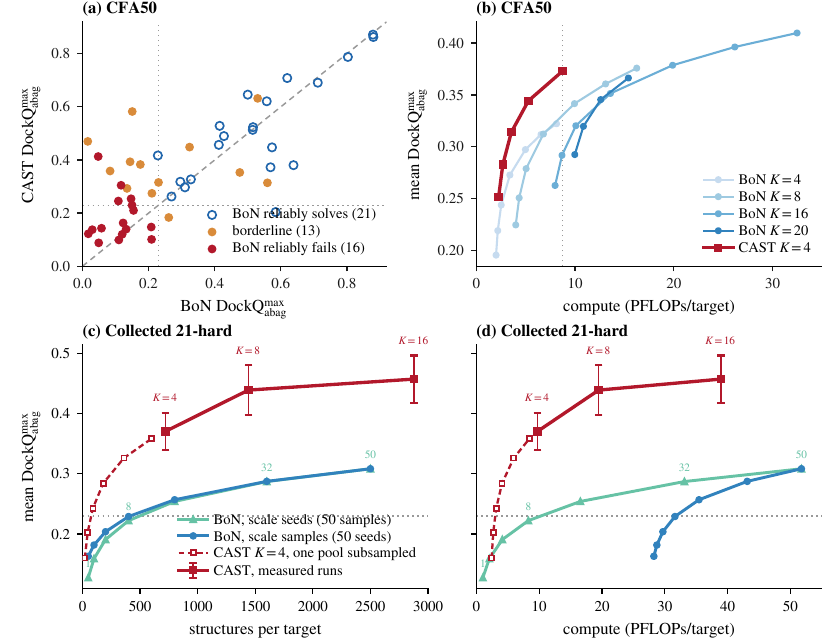}
\end{center}
\caption{\textbf{(a)} Per-target $\mathrm{DockQ}^{\max}_{\mathrm{abag}}$ on CFA50, both arms at
$720$ structures and the same $4$ alignment subsamples, BoN spending it as $4\times180$. Points
above the dashed diagonal are CAST wins, dotted lines mark the CAPRI-acceptable threshold $0.23$,
and fill encodes the grouping of Table~\ref{tab:main}. \textbf{(b)} The same set against compute,
BoN drawn separately at each conditioning count $K$. \textbf{(c)} and \textbf{(d)} show the
collected 21-hard targets, against structures
and against compute. Best-of-$N$ appears twice, scaling conditionings at $50$ samples each and scaling
samples at $50$ conditionings. Every Best-of-$N$
point in (b), (c) and (d) is the exact expectation over subsets of that size (\S\ref{app:expect}).
Filled squares are separate CAST runs at $K=4$, $8$ and $16$ along $N=180K$,
labelled by $K$ and with bars giving the standard error over the $21$ targets. Open squares on the
dashed line, and CAST's points below $720$ in (b), are its single $K=4$ pool of $720$ drawn at fewer
particles.}
\label{fig:main4}
\end{figure}

\textbf{At a matched budget CAST solves $10$ more targets.} Table~\ref{tab:main} and
Figure~\ref{fig:main4}(a) give the comparison. Means are $0.373$ against $0.322$, a difference of
$+0.051$ ($+15.8\%$), with $31$ wins out of $50$ at Wilcoxon signed-rank $p=0.012$, and $10$ extra
targets reaching $0.23$, $36$ against $26$. On the $16$ targets BoN \emph{reliably fails}, CAST wins
$12$ at $+0.071$ ($+63.9\%$) and reaches $0.23$ on $4$ against $0$, which BoN does under no
$4$-conditioning draw. On the $13$ \emph{borderline} targets it wins $10$ at $+0.131$ ($+51.8\%$)
and reaches $0.23$ on $12$ against $6$. On the $21$ BoN \emph{reliably solves} the difference is
$-0.015$ ($-2.8\%$), which the Wilcoxon test cannot separate from zero ($p=0.99$).

\begin{table}[!ht]
\caption{$\mathrm{DockQ}^{\max}_{\mathrm{abag}}$ over the $50$ CFA50 targets for CAST and
BoN, both at a \emph{matched} budget of $720$ structures and the same $4$ conditionings,
BoN spending it as $4\times180$. Rows are grouped by how reliably the
\emph{BoN alone} solves a target (\S\ref{app:group}): \emph{reliably fails}
is $\le10\%$ of draws reaching $\mathrm{DockQ}^{\max}_{\mathrm{abag}}\ge0.23$, \emph{reliably solves} is
$\ge90\%$, and \emph{borderline} is the rest, sizes in brackets. SITE measures how close an arm 
gets to the true site, using the centroid distance between the predicted and true epitopes for its closest structure. 
Lower values are better. REACH $0.23$ counts targets on
which that arm attains $\mathrm{DockQ}^{\max}_{\mathrm{abag}}\ge0.23$, CAST WINS is a paired
count spanning both rows, and $p$ is a two-sided Wilcoxon signed-rank test on the per-target
differences within that subset.}
\label{tab:main}
\begin{center}
\small
\setlength{\tabcolsep}{3.5pt}
\begin{tabular}{llrrrrr}
\toprule
\multicolumn{1}{c}{\bf SUBSET} &\multicolumn{1}{c}{\bf ARM}
&\multicolumn{1}{c}{\bf MEAN $\mathrm{DockQ}^{\max}_{\mathrm{abag}}$}
&\multicolumn{1}{c}{\bf SITE (\AA{})}
&\multicolumn{1}{c}{\bf REACH $0.23$}
&\multicolumn{1}{c}{\bf CAST WINS}
&\multicolumn{1}{c}{\bf $p$} \\
\midrule
\multirow{2}{*}{All targets ($50$)}
  & CAST & ${\bf 0.373}$ & ${\bf 1.75}$ & ${\bf 36}$ & \multirow{2}{*}{$31$} & \multirow{2}{*}{$0.012$} \\
  & BoN  & $0.322$ & $2.96$ & $26$ & & \\
\addlinespace
\multirow{2}{*}{BoN reliably fails ($16$)}
  & CAST & ${\bf 0.182}$ & ${\bf 3.02}$ & ${\bf 4}$ & \multirow{2}{*}{$12$} & \multirow{2}{*}{$0.020$} \\
  & BoN  & $0.111$ & $5.52$ & $0$ & & \\
\addlinespace
\multirow{2}{*}{Borderline ($13$)}
  & CAST & ${\bf 0.385}$ & ${\bf 1.67}$ & ${\bf 12}$ & \multirow{2}{*}{$10$} & \multirow{2}{*}{$0.040$} \\
  & BoN  & $0.253$ & $3.41$ & $6$ & & \\
\addlinespace
\multirow{2}{*}{BoN reliably solves ($21$)}
  & CAST & $0.511$ & $0.83$ & $20$ & \multirow{2}{*}{$9$} & \multirow{2}{*}{$0.99$} \\
  & BoN  & ${\bf 0.525}$ & ${\bf 0.72}$ & $20$ & & \\
\bottomrule
\end{tabular}
\end{center}
\end{table}

\textbf{CAST reaches binding sites the base model rarely samples.} The mechanism is budget
reallocation, by construction. CAST stratifies its pool over the $U\times S=60$ cells
and so spends most of it on wrong sites, while BoN spends everything where the
base model already goes (SITE, Table~\ref{tab:main}). On the $16$ targets
BoN reliably fails, its closest structure over the whole
$720$ sits $5.52$\,\AA{} from the true epitope on average against CAST's $3.02$\,\AA{}, and on $4$ of
them CAST lands within $6$\,\AA{} where no structure in BoN's pool does. $2$ more sit in the
borderline group, and where BoN already solves there is no difference, $0.83$\,\AA{} against
$0.72$\,\AA{}. Every candidate site receives particles whatever the base model prefers, so CAST has
the chance to explore the right site even where the base model puts little mass, which BoN at the
same budget does not.

\textbf{CAST at $K=4$ matches BoN at $K=20$ from $5\times$ fewer conditionings and $1.5\times$ less
compute.} Holding the budget at $720$ structures, we vary only how it splits between conditionings
and samples. BoN improves with conditioning count, from $0.322$ at $K=4$ to $0.366$ at $K=20$. CAST returns $0.373$ from $K=4$, within $+0.007$ of the top rung (Wilcoxon $p=0.77$), which the test
does not separate. Cost is what
separates them, since that rung pays $20$ trunk passes and $15.40$ PFLOPs against CAST's $8.72$
(Figure~\ref{fig:main4}(b), \S\ref{app:alloc}). Charging the map's $150$-structure pilot pool takes
CAST to $10.04$. CAST therefore reaches the same level from $5\times$ fewer
conditionings and $1.5\times$ less compute.

\textbf{On collected 21-hard CAST leads at every budget.}
We vary the budget on collected 21-hard. BoN can spend more two ways, by scaling
conditionings or by scaling samples (Figure~\ref{fig:main4}(c,d)). The two axes agree in structures
and diverge in compute. Both baseline curves keep climbing and flatten as they go, so each added
structure buys less than the last, and neither reaches $0.31$ at any budget we ran. CAST passes that
level at $720$ structures and $9.73$ PFLOPs. At that matched $720$ it brings $18$ of the $21$ to
CAPRI-acceptable DockQ against BoN's $9$, and no BoN allocation of the same budget passes $11$. The baseline's whole $2500$-structure pool costs
$5.3\times$ that compute and still sits $0.062$ below, a $20.1\%$ lead for CAST. Scaling CAST further needs both of its axes at once. We raise samples and latents together along
$N=180K$, which holds each latent's share of every cell fixed. Each rung is a separate run, reaching
$0.439$ at $K=8$ and $0.457$ at $K=16$. The lead over the best baseline widens from $0.062$ to $0.149$, a
$48.3\%$ gain, while still spending three quarters of compute.

\textbf{The gain transfers to a second backbone.} We repeat the matched-budget and scaling comparisons on
OpenDDE \citep{opendde2026}, with no hyperparameter retuned. Over $40$ difficulty-balanced targets CAST reaches $0.454$
against BoN's $0.372$ ($p<10^{-4}$), bringing $27$ to an acceptable pose against $20$. The gain
again sits where BoN is unreliable, $+0.111$ on the $16$ it reliably fails, where CAST also outscales BoN and exceeds every allocation of the budget (\S\ref{app:opendde}).  

\subsection{Steered baselines and ablations}
\label{sec:res-baselines}

FK, NS, BP and ReD (\S\ref{sec:experiments}) are four other ways to spend the same budget on steering,
by resampling without a gradient, by searching over the initial noise, by perturbing the
conditioning with no objective, and by repelling from the structures already produced. All four run
at the same $K=4$ conditionings and the same
$4\times180$ allocation as CAST and BoN, so every arm in Table~\ref{tab:baselines} sees $720$
structures at the same compute to within $7\%$.

\textbf{Only CAST converts more failed targets.} CAST reaches $36$ of the $50$ targets, against $30$
for FK, $27$ for ReD, $26$ for BoN, $24$ for BP and $22$ for NS. Exact McNemar on the paired threshold crossings
gives $p=0.11$, $0.012$, $0.013$, $0.004$ and $<10^{-3}$ respectively. The separation is sharpest on the
$16$ BoN reliably fails, where CAST reaches $0.182$ and brings $4$ to $0.23$, against ReD's
$0.131$ and $1$, FK's $0.119$ and $1$, and none for NS and BP. CAST wins $13$ of $16$
over ReD ($p=0.025$). What separates them is
transport. FK resamples on the same site reward, so it can only keep
particles that already drift toward a candidate. CAST's gradient moves them there.
Where BoN already solves, FK is best at $0.540$, since it uses one arrival radius to
CAST's four, so it holds $48$ particles per candidate against $12$, and concentration wins where the
site is already found. The two cancel, so CAST's lead over FK across all $50$ is $+0.036$ with a
bootstrap $95\%$ interval of $[-0.003,\,0.079]$ ($p=0.104$), while on
the $29$ BoN does not reliably solve it is $+0.082$ ($p=0.002$), reaching $16$ against $9$.

\textbf{Diversity alone does not explain the gain.} BP perturbs the conditioning with no objective,
and lands at $0.313$ against BoN's $0.322$. ReD, which spreads structures apart by repelling from
earlier ones, lands at $0.311$. Neither is separable from the unsteered control over the $50$ paired
targets, at $p=0.92$ and $p=0.48$, so variety on its own does not account for the gain. NS is
weakest of all at $0.238$.
Searching the initial noise lets it score only finished structures, and by then the site is
decided, so a site reward can no longer tell candidates apart. It is at its maximum by the first
generation on $46$ of the $50$ targets, and the rest of the search shrinks the pool without improving
it. Over all $50$ targets CAST has the highest mean, $0.373$ against $0.337$, $0.313$, $0.311$ and
$0.238$, gains of $10.7\%$, $19.2\%$, $19.9\%$ and $56.7\%$, and reaches $0.23$ on $36$ targets
against $30$, $24$, $27$ and $22$.

\begin{table}[!ht]
\caption{Steered baselines at a matched $720$ structures, against CAST and BoN on CFA50. Mean
$\mathrm{DockQ}^{\max}_{\mathrm{abag}}$ over all $50$ targets and within each group of
Table~\ref{tab:main}, and the number of targets reaching $0.23$. Groups are named by BoN, so the
column headed \emph{fails} is the subset BoN reliably fails. $p$ is a two-sided Wilcoxon
signed-rank test of CAST against that arm on per-target differences.}
\label{tab:baselines}
\begin{center}
\small
\setlength{\tabcolsep}{4.5pt}
\begin{tabular}{lrrrrrrr}
\toprule
\multicolumn{1}{c}{\bf ARM} &\multicolumn{1}{c}{\bf PFLOPs/TGT}
&\multicolumn{4}{c}{\bf MEAN $\mathrm{DockQ}^{\max}_{\mathrm{abag}}$}
&\multicolumn{1}{c}{\bf REACH $0.23$} &\multicolumn{1}{c}{\bf $p$} \\
 & &\multicolumn{1}{c}{all ($50$)} &\multicolumn{1}{c}{fails ($16$)}
&\multicolumn{1}{c}{border ($13$)} &\multicolumn{1}{c}{solves ($21$)} & & \\
\midrule
CAST & $8.72$ & ${\bf 0.373}$ & ${\bf 0.182}$ & ${\bf 0.385}$ & $0.511$ & ${\bf 36}$ & \\
BoN  & $8.12$ & $0.322$ & $0.111$ & $0.253$ & $0.525$ & $26$ & $0.012$ \\
\addlinespace
FK   & $8.69$ & $0.337$ & $0.119$ & $0.279$ & ${\bf 0.540}$ & $30$ & $0.104$ \\
NS   & $8.12$ & $0.238$ & $0.076$ & $0.180$ & $0.398$ & $22$ & $<10^{-9}$ \\
BP   & $8.12$ & $0.313$ & $0.103$ & $0.222$ & $0.529$ & $24$ & $0.013$ \\
ReD  & $8.12$ & $0.311$ & $0.131$ & $0.262$ & $0.478$ & $27$ & $0.003$ \\
\bottomrule
\end{tabular}
\end{center}
\end{table}

\textbf{Ablations.} Six ablations each change one field of the reference arm at $K=4$ and $N=720$,
over all $50$ CFA50 targets (\S\ref{app:abl}, Table~\ref{tab:abl}). Three carry the method. Removing
the gradient costs $0.063$, shuffling the map $0.054$ and collapsing the umbrella sweep $0.012$,
so the gain is the tilt and the site it names. A separate run per conditioning moves the mean by only $0.006$, and its spread across
conditionings is $\pm0.075$, so the pool buys certainty and cost, not accuracy. Resampling instead trades
between the groups, gaining $+0.034$ on borderline and paying $0.039$ on
fails, with no change in the pooled mean. CAST restricts resampling to two early events to limit pool collapse. An oracle map adds $0.111$ and takes reach from $36$ to $42$, the headroom from better maps.

\section{Discussion and conclusion}
\label{sec:discussion}

CAST carries a site hypothesis and its conditioning latent in the particle state, so many hypotheses
and $K$ subsamples share one pool. On
CFA50 at a matched $720$ structures it brings $36$ of $50$ targets to an acceptable pose against
$26$, and raises mean $\mathrm{DockQ}^{\max}_{\mathrm{abag}}$ from $0.322$ to $0.373$, a gain of
$15.8\%$. That gain sits on the
$29$ targets BoN does not reliably solve, and it matches BoN's best allocation of the budget from
$1.5\times$ less compute. On collected 21-hard it brings $18$ of the $21$ to threshold
against $9$, where BoN's largest pool spends $5.3\times$ the compute and still trails by
$20.1\%$. Raising latents and samples together carries it to $0.457$, where Best-of-$N$ flattens. Both datasets are selected using the unsteered model alone, CFA50 on a
$200$-structure pool and collected 21-hard on a $1{,}000$-seed reference, so difficulty here is a
property of the base model.

Two parts of this generalise beyond CAST. The mechanism is a site-measurable reward and a twisted
sampler, so any co-folding model with a diffusion decoder can carry the same objective, which the
OpenDDE transfer confirms. The pose-invariance guarantee
comes from that class of reward, so any site variable on the interface
separates exploring sites from selecting poses at one.

CAST does not beat BoN on the targets BoN already solves, where the difference is not separable from
zero. This trade-off is by construction. CAST stratifies its pool over the candidate
cells and so spends most of it on wrong sites, while BoN spends everything where the base model
already goes. Exploring sites is paid for out of the budget that would otherwise exploit one. The
pose-invariance result then bounds what steering can reach. CAST moves the antibody to a site and
leaves everything after that to the base model, so it cannot correct a pose the base model gets
wrong at the true site.

CAST addresses one of co-folding's two failures. It improves site coverage, and what the pool
contains bounds the result before ranking. Selection sits one stage beyond, and ranking a pool
without a reference remains open. CAST runs on a frozen model and takes a site as an argument, so a
hypothesis from an epitope predictor or an experiment turns directly into structures at that site.
It needs no training and no reference structure, so testing candidate epitopes costs only a single
sampling run. Better epitope prediction and ranking each improve what it returns without changing
the method.

\bibliographystyle{plainnat}
\bibliography{cast}

%\clearpage
\appendix

\section{Extended related work}
\label{sec:related}

Two axes organise the work we build on. The first is \emph{what} a method diversifies, the
\textsc{pose} a molecule adopts, meaning its orientation and internal conformation, or the
\textsc{site} on a partner at which it binds. The second is \emph{where} the intervention enters,
the trunk that produces the conditioning representation, which we call \emph{conditioning-side} and
which includes restraints, or the diffusion loop that consumes it. Pose work is crowded into both,
whereas site-directed work has so far intervened at the model input. Cutting across both axes,
inference-time scaling is usually cast as search, a sampler proposing candidates and a verifier
ranking them \citep{ma2025inference,zhang2025classical,jajal2025evo}. CAST changes what the sampler
proposes by steering and leaves the ranking to a separate verifier (\S\ref{sec:discussion}).

\textbf{Pose, conditioning-side.} These methods perturb the conditioning and let a frozen decoder
follow. Masking or subsampling the alignment weakens the coevolutionary signal that pins the model
to one state \citep{afsample2,afsample3}. Rescaling the pair latent by a scalar does the same one
layer later \citep{boltzsample2026}, and trained channel-wise affine transforms of the
pre-Pairformer latent learn a reusable direction instead of a random one \citep{confornets}. The
first two are gradient-free, which is why alignment subsampling supplies the latents CAST amortises.
None of the three can \emph{request} a named change at inference. Two perturb without a destination,
and the third fixes its direction before sampling begins.

\textbf{Pose, diffusion-side.} Here the intervention enters the sampler. One group aims at a target
it can trust. \citet{embedopt} fits a conditional embedding to experimental measurements, so the
data fixes the state it is steering to. \citet{forceguided} guides with a learned force network
carrying physical prior knowledge, so it aims at an equilibrium ensemble rather than one named
state. Both targets are correct by construction in a way a predicted epitope is not, and neither of
them is a location on the antigen.
A second group needs no target at all. Particles are coupled so that they repel, on
coordinates \citep{corso2024particle,af3red2025} or on trunk embeddings \citep{conforflux2026}.
Elsewhere, trajectories are driven \emph{away} from the model's own default prediction, which
distorts the marginal sampled. Both buy spread without aim. \citet{conformix2025} is the closest
work in method and the furthest in aim. It runs the twisted sequential-Monte-Carlo scheme of
\citet{wu2023practical} on a co-folding model, under RMSD potentials that repel from the default
structure. Retraining under flow matching \citep{alphaflow} moves the same axis at a training cost
we do not pay. CAST also couples particles, but its reward reads the site alone, so the pose at a
site is left to the base model (Proposition~\ref{prop:invariance}), where an RMSD
potential moves it.

\textbf{Pose diversity is not site diversity.} The potentials above are RMSD collective variables
read off a superposed structure, Kabsch-aligned and masked to secondary structure in
\citet{conformix2025}, superposed C$\alpha$ against earlier seeds in \citet{af3red2025}. Such a
variable rewards any departure from the reference, so the mass it moves follows the base model's own
ranking of the alternatives. A different site can come out of that, but the potential never chooses
which, and it cannot be pointed at a particular epitope. Naming an epitope needs a
variable that reads interface placement with the antigen held fixed, which is the site reward of
\S\ref{sec:cast}.

\textbf{Site, conditioning-side.} The closest neighbour supplies the site directly.
\citet{clifford2026pocket} turn a B-cell epitope prediction into a soft pocket restraint on Chai-1
\citep{chai1} and report better accuracy and diversity. The restraint is an input, fixed before the
trunk runs, so each site hypothesis needs its own run. Protenix-v2 \citep{protenixv2} exposes the same interface but ships
no weights for its pocket embedder. A restraint is only as good as the epitope it encodes, and
prediction is open \citep{liu2024asep}. Boltz-Perturb instead injects time-annealed
noise into the denoiser's conditioning \citep{boltzperturb2025}. The noise lands on the per-token
representation, the pair representation, or both, masked in the source to the pairs touching the
ligand. Boltz-Perturb is training-free and establishes that the correct binding-mode basins already
exist inside a frozen model, but it was built for small-molecule binding modes and the perturbation
carries no objective.

\textbf{Site, diffusion-side.} What exists here is a sampler, not a destination. Twisted Diffusion
Sampling steers a diffusion model toward a conditional target with a gradient-guided proposal,
importance weights and resampling, and is asymptotically exact \citep{wu2023practical}. Feynman--Kac
steering reweights trajectories from potentials alone \citep{singhal2025fk}, and has since been
carried to protein design \citep{hartman2025fkprotein}. Both are backbone-agnostic and
training-free, and both prescribe \emph{how} to steer, not \emph{what} to steer toward.

\textbf{Site, both sides.} ACER \citep{acer2026} reaches our diagnosis independently in
protein--ligand co-folding, that a frozen model drops the ligand into its dominant pocket and
generalises poorly to allosteric or novel sites, and is likewise training-free and reference-free.
Decoy copies of the ligand enter the trunk input to occlude that pocket and are discarded before
denoising, one Pairformer pass per configuration. A Feynman--Kac potential \citep{singhal2025fk}
then repels the ligand from pockets occupied in earlier rounds, each anchor read off the previous
round's pose, so the rounds cannot be batched. A second stage turns the pockets found into
restraints and ranks fresh samples drawn under them by ensemble-weighted confidence. Every stage is
exclusionary or post hoc, so no site is named in advance. CAST instead attracts toward a named
candidate, reads the site alone, and sweeps candidates inside one pool at $\mathcal{O}(N+K)$.

\section{Method details}
\label{app:method}

% Unlimited pages allowed here, after the bibliography, but reviewers are NOT required to read the
% appendix, so anything load-bearing must stay in the main text.

\subsection{The sampler in full}
\label{app:algorithm}

\begin{algorithm}[!ht]
\caption{CAST. Every line acts on each particle, whose index is suppressed. Each particle carries a
latent $k$ and a cell $b$. The guided step aims at that cell, the weight corrects the reward level
between scoring events, and resampling is confined within cells.}
\label{alg:cast}
\begin{algorithmic}[1]
\State \textbf{input} latents $z_1,\dots,z_K$, candidates $\{\mu_s\}$, radii $\{d^\star_u\}$, pool
size $N$, window $\mathcal{W}$
\State give each particle a latent $k$ and a cell $b$, spread evenly over
$\mathcal{U}\times\mathcal{S}$ (\S\ref{app:assign})
\State $x_T\sim\normal(0,\sigma_T^2\I)$, \;\; $\log w\gets 0$, \;\; $g_T\gets 0$
\For{$t=T,\dots,1$}
  \State recenter and randomly rotate $x_t$, carrying $g_t$ through the same transform
  \If{$\beta_t>0$}  \Comment{the sampler stops churning at low $\sigma_t$}
    \State $\varepsilon\sim\normal(0,\I)$, \;\; $\tilde x_t\gets x_t+g_t+\beta_t\varepsilon$, \;\;
           $\log w \mathrel{+}= -\big(\norm{g_t}^2+2\beta_t\langle\varepsilon,g_t\rangle\big)
           /(2\beta_t^2)$ \Comment{\eqref{eq:ratios}}
  \Else
    \State $\tilde x_t\gets x_t$
  \EndIf
  \State $\hat x_0\gets D_\theta(\tilde x_t,\hat\sigma_t\g z_k)$, \;\;
         $x_{t-1}\gets\tilde x_t+(\sigma_{t-1}-\hat\sigma_t)(\tilde x_t-\hat x_0)/\hat\sigma_t$
  \State $g_{t-1}\gets\mathrm{clip}\big[\eta_{t-1}\,\nabla_{\hat x_0} r_b(\hat x_0)\big]$
         \Comment{reused next step, \S\ref{app:guidance}}
  \If{$t$ is scoring event $m$}
    \State $\bar x_0\gets$ deterministic $4$-step Euler rollout of $\tilde x_t$ down the remaining
           $\sigma$ \Comment{$4$ denoiser passes, \S\ref{app:compute}}
    \State $r_m\gets r_b(\bar x_0)$, \;\;
           $\log w \mathrel{+}= \lambda(r_m-r_{m-1})$ \Comment{$r_{-1}\!:=\!r_0$}
    \If{first event with non-zero reward spread}
      \State solve $\lambda$ for a target ESS of $0.4$, capped, then hold it, \;\;
             $\log w \mathrel{+}= \lambda r_0$ \Comment{level correction, \S\ref{app:amortisation}}
    \EndIf
    \State \textbf{if} $t\in\mathcal{W}$: resample each cell's $N_b$ particles among themselves when
           $\mathrm{ESS}_b<\gamma N_b$, at $\gamma=0.5$
  \EndIf
\EndFor
\State \textbf{return} the $N$ structures $x_0$, ranked by confidence
\end{algorithmic}
\end{algorithm}

The sampler reads two different estimates of $x_0$. Guidance uses the one-step $\hat x_0$ the
reverse step already produced, which is free. The reward at a scoring event uses $\bar x_0$, a
deterministic $4$-step Euler rollout from the current state to $\sigma\approx0$ with no churn,
because a one-step estimate at mid-$\sigma$ is still diffuse enough to blur the contact set the
reward reads. Eight events at four passes each are the $32$ extra denoiser passes charged in
\S\ref{app:compute}.

At a resampling event the weights reset, and both the difference baseline $r^i_{m-1}$ and
the pending displacement are reindexed by ancestry, since a clone continues its ancestor's trajectory
and not that of the particle previously in its slot.

The window $\mathcal{W}$ sits early because the site is decided early. Once a particle's site is
fixed, resampling can only duplicate or discard particles, not redirect them. By
Proposition~\ref{prop:invariance} the reward also says nothing about the pose, which is the only
thing still free. Resampling late would therefore replace distinct structures with copies of one
another, spending pose diversity and buying nothing.

Three approximations separate this sampler from the ideal target of \S\ref{sec:theory}, and they are
not of the same kind. The plug-in potential and the straight-through Jacobian both act through the
displacement $g^i_t$, which moves the mean of the Gaussian injection and nothing else. Whatever
displacement they produce, \eqref{eq:ratios} is the exact likelihood ratio for that injection, so the
weights price it in full. A poor $g^i_t$ therefore costs proposal quality, and so efficiency, while
leaving the target alone. The window is the one approximation that costs exactness rather than efficiency.

We apply weights only inside $\mathcal{W}$ and return the ensemble unweighted, which forfeits the
asymptotic exactness of \citet{wu2023practical}, since that holds only if every weight is carried to
the end. Exactness would buy a correctly weighted distribution over the pool, and we never use one.
Every number we report is $\mathrm{DockQ}^{\max}$, the best single structure the pool contains. A
maximum depends only on whether the pool reaches the right region, not on how much probability mass
sits on each member of it.

At the deployed schedule, scoring events fall every $20$ of $200$ steps from $0.25T$, and
$\mathcal{W}=[0.25T,0.35T]$ admits two. The first cannot be tilted, because the difference potential
sets the baseline $r^i_0$ there, so its spread is zero, $\lambda$ does not calibrate, and the reward
increment and level correction are both skipped. It still resamples on the injection ratio of
\eqref{eq:ratios} alone, which selects mildly against the particles guidance moved furthest.
$\lambda$ is solved at the second event and then held, so that is the one tilted resample of the run.
The tilt itself is carried continuously, the displacement being applied at $159$ of $200$ steps.
Only the steps inside $\mathcal{W}$ carry a weight against it, so the remainder is an unweighted
heuristic drift rather than part of the sequential correction.

\subsection{Proofs}
\label{app:proofs}

\begin{proof}[Proof of Proposition~\ref{prop:invariance}]
Fix $k$, and write $Z=\sum_k\rho_k\int p(a\g z_k)e^{\lambda r(a)}\,\mathrm{d}a$ for the
normalising constant of \eqref{eq:tilt}. By \eqref{eq:factor} and \eqref{eq:tilt},
$q(k,a,R,c)=Z^{-1}\rho_k\,p(a\g z_k)\,p(R,c\g a,z_k)\,e^{\lambda r(a)}$. Marginalising the pose
leaves $q(k,a)=Z^{-1}\rho_k\,p(a\g z_k)\,e^{\lambda r(a)}$, because $p(R,c\g a,z_k)$ is a normalised
conditional and the tilt does not depend on $(R,c)$. Dividing gives the first claim. For the second,
$q(k\g a)\propto\rho_k\,p(a\g z_k)\,e^{\lambda r(a)}$, and the tilt is a function of $a$ alone, so it
cancels in the normalisation over $k$ and $q(k\g a)=p(k\g a)$. Hence
$q(R,c\g a)=\sum_k q(k\g a)\,p(R,c\g a,z_k)=\sum_k p(k\g a)\,p(R,c\g a,z_k)=p(R,c\g a)$.
\end{proof}

\begin{proof}[Proof of Corollary~\ref{cor:ball}]
The hinged reward $r_b$ is zero on $\mathcal{B}_b$ and strictly negative off it, so
$e^{\lambda r_b}\to\mathbf{1}\{a\in\mathcal{B}_b\}$ pointwise as $\lambda\to\infty$. Since $r_b\le0$
the family is bounded by $1$, so dominated convergence carries the limit through both the product and
its normaliser, and the normaliser tends to $p_{\mathrm{mix}}(a\in\mathcal{B}_b)$, which is positive
by assumption. Normalising the
product of that limit with $p_{\mathrm{mix}}$, or with $p_\theta(\cdot\g z_k)$ for the per-latent
statement, gives the conditioning. The event is measurable with respect to $a$ by
Definition~\ref{def:site}. Without that positivity the limit is a collapse, not a
conditioning, since the tilt would then concentrate on a set the base mixture never visits.
\end{proof}

Proposition~\ref{prop:invariance} constrains $q_b$, a distribution over finished structures, not the
path to one. No step along that path is pose-preserving, and none should be. The pose conditional
$p(R,c\g a)$ depends on $a$, so a particle moving to a new site must adapt its pose to land on a
structure the model recognises. A rigid translation would carry the old site's pose to the new one.
Since $r_b$ reads $a$ only through the separation $a-\mu_s$, a difference of the paratope centroid
and the candidate centroid, $\nabla_{\hat x_0}r_b$ is non-zero only on the paratope
and candidate-epitope atoms, so a step stretches each chain instead of moving it rigidly. The
proposition asks nothing of the proposal, and the denoiser repairs the stretch at the next step.

\subsection{Consequences of sharing one pool}
\label{app:amortisation}

The potential ratio of \eqref{eq:ratios} is taken in \emph{difference} form, each event contributing
$\exp(\lambda(r^i_m-r^i_{m-1}))$, because the absolute reward is miscalibrated at high noise where
$\hat x_0$ is a blurred average. At constant $\lambda$ the increments telescope to
$\exp(\lambda(r^i_M-r^i_0))$ not to $\exp(\lambda r^i_M)$, dividing out each particle's own
starting level.

Across multiple latents in one pool, $r^i_0$ records where the trunk under
$z_{k(i)}$ already places the antibody, and dividing it out leaves the cross-latent comparison at chance,
with pairwise AUC $0.484$ to $0.510$ over the eight scoring events. Adding $\lambda r^i_0$ back once
restores the level. At $K=1$ there is no cross-latent comparison to restore, so the correction has nothing to act on.
It is a joint-pool setting rather than a free parameter.

\subsection{Latent assignment and the cell budget}
\label{app:assign}

For particle index $i$, with $U=|\mathcal{U}|$ umbrellas and $S=|\mathcal{S}|$ sites, the cell and
the latent are
\begin{equation*}
b(i)=u(i)\,S+s(i),\quad u(i)=i\bmod U,\quad s(i)=\lfloor i/U\rfloor\bmod S,
\qquad
k(i)=\big\lfloor i/(US)\big\rfloor \bmod K,
\end{equation*}
a full cross-product of distance against site, so every cell receives all $K$ latents. Umbrella ids
are read from the $d^\star$ each particle actually carries rather than from its position, so a sweep
repeating a value groups by that value. For a sweep of distinct $d^\star$ this is the $i\bmod U$
above up to relabelling, and only the partition matters downstream. The split is exactly even when $K$ divides the repeat count
$N/|\mathcal{U}\times\mathcal{S}|$, which holds at the budget we report, $12$ repeats over $60$ cells
at $K=4$. Otherwise the first few latents take one extra repeat each, and a comparison across $K$ is
confounded with that imbalance.

Two floors bound $K$. A cell needs about six particles to resample at all, measured at four particles
over four umbrellas where the ancestors came back unpermuted at every event, and about three per
latent, or one resample can drop a latent by chance. At our budget a cell holds $12$, which clears
the first floor and allows $12/3=4$ latents, the $K=4$ we report.

\subsection{Guidance schedule and cap}
\label{app:guidance}

The gradient is taken with respect to the denoised \emph{coordinates} of a reward that reads only the
site, so it is supported on the paratope and candidate-epitope atoms, with equal and opposite
coefficients that draw the two centroids together. Its step is
\begin{equation*}
\eta_t=\kappa\,\frac{\sigma_t^2}{\sqrt{1+\sigma_t^2}},
\end{equation*}
at $\kappa=75$, the variance-exploding image of the twist coefficient $(1-\bar\alpha_t)/\sqrt{\bar\alpha_t}$ that
\citet{wu2023practical} use on a variance-preserving schedule with signal retention $\bar\alpha_t$.
It grows with the noise level, so the permitted move is large while the endpoint is still undecided
and vanishes as the structure commits, and it is therefore fixed by the schedule and never tuned. The constant
$\kappa$ is the proposal's own strength and is set independently of the potential's $\lambda$, so the
displacement need not equal the tilted score of \eqref{eq:plugin}. Nothing is lost by that, since
\eqref{eq:ratios} is exact for whichever displacement is applied, so a mismatch costs proposal
quality while leaving correctness intact. The
clip is a per-particle cap of $6$\,\AA{}, expressed as a root-mean-square displacement per coordinate, so the same
setting transfers across targets of different size. The displacement is carried through each step's rigid
augmentation, since $r$ is rotation-invariant but $\nabla r$ is a vector field and a shift computed in
the previous step's frame would point the wrong way after a fresh random rotation.

\subsection{What the umbrella sweep corrects}
\label{app:umbrella}

With the Huber knee of $\ell_\delta$ at $\delta=4$\,\AA{}, the reward is linear, not quadratic
across most of the pool. Over all cell evaluations of a completed run $94.4\%$ are active
($d>d^\star$), and $93.4\%$ of those lie beyond $\delta$, where the gradient magnitude is the constant
$2\delta$. Every active particle therefore receives the same push magnitude whatever its distance, so
the transport carries direction but no distance information, and a single $d^\star$ cannot be
corrected by the pull weakening near arrival. Sweeping $u$ supplies that distance by construction.
This predicts a refinement, and A3 (Table~\ref{tab:abl}) matches that shape.
The gain is largest where BoN already solves, $+0.021$, and smallest on the targets it fails,
$+0.004$, so the sweep adjusts an arrival distance once the site is roughly right.

\subsection{Building the candidate map}
\label{app:map}

\paragraph{Pool modes.} For each structure in an unsteered pool we superpose the antigen onto a
common reference. A contact residue is one whose C$\alpha$ lies within $10$\,\AA{} of any antibody
C$\alpha$, and each structure contributes the centroid of its own contacts. Modes are then found
greedily on that cloud. We seed at the centroid with the most neighbours inside one $12$\,\AA{}
epitope footprint, claim every centroid within it, and repeat. A mode's patch is the set of residues
contacted by at least half of the structures assigned to it, and its score is its population share.
We keep the top three, since the first carries most of the pool mass but is not reliably the right
site, and the second and third recover many of the targets where it is wrong. Each mode takes its
consensus independently, instead of assigning every residue to its strongest mode. That matters
because modes overlap at their rims, and stripping a shared rim shifts the weaker centroid off the
site.

\paragraph{Surface tiles.} Tiles come from a farthest-point covering of the antigen. We seed at the
residue farthest from the antigen centroid, claim everything unclaimed within a radius, and repeat.
That radius is binary-searched to a fixed tile count and is not fixed in advance. A fixed radius
makes the tile count scale with antigen size, so a large antigen produces far more tiles than the
candidate budget $S$ can hold, and truncating drops most of its surface. Fixing the count instead
keeps the budget per target constant, and simply makes the tiles coarser on a large antigen. No
surface filter is applied, since dropping the inner shell assumes the epitope is exposed, and
groove-binding epitopes would then fall in no tile at all.

\paragraph{The hybrid.} Modes are listed first and kept verbatim, because they come from real poses
and are better centred, whereas tiles are comprehensive but arbitrarily centred. Every tile is then
appended, ranked by the fraction of its residues carrying mode support. That ranking uses only the
pool's own mode mass, so it is ground-truth free, and it concentrates budget sharply because most
tiles have no mode support at all. Ties break by a stable sort on the tiling's own order. That
matters because the map is truncated to the candidate budget $S$, so an arbitrary tie-break would
silently change which sites receive particles. Patches are finally de-duplicated, and any patch of
fewer than $3$ residues is dropped, since a smaller one has no meaningful centroid. At a $6$\,\AA{}
footprint the hybrid then reaches more targets than either ingredient alone.

\section{Experimental details}
\label{app:experiments}

\subsection{Compute accounting}
\label{app:compute}

Deferring the decoder to the $K$-th subsample leaves the first $K-1$ to return one unsteered draw
each, so a joint run emits $N+K-1$ structures while its pool is $N$. We exclude those $K-1$ from
every metric and still charge all $K$ trunk passes, so the method pays for them in cost and collects
nothing from them in quality. The exclusion changes no reported number. Across $71$ CAST units no
such draw has ever held its unit's maximum, and the closest any came was $0.014$ below it.

Reported FLOPs are per target, averaged over all targets, and three terms enter the count.
The \textbf{trunk} is charged once per conditioning and stays outside the structure count, which is
why arms at the same $720$ structures differ in compute, $8.12$ PFLOPs at $K=4$ against $15.40$ at
$K=20$. The \textbf{denoiser} is charged per structure, CAST's at its measured $1.16\times$ a BoN
structure's, $232$ passes against $200$, the excess being the mini-rollout at each scoring event, so
the gradient CAST adds is paid for and never assumed free. The \textbf{confidence head} is charged
once per finished structure and every arm pays it alike.

\textbf{The pilot pool the map needs.} The map's modes are read from an unsteered pool, which a new
target must have or pay for. Ours are built from $150$ structures drawn from
an unsteered BoN pool. Every CAST figure we report is the steered run alone, on targets whose pilot pool
already existed and was reused. A fresh target pays for those $150$ structures at the BoN
per-structure rate of $0.0088$ PFLOPs, so a fixed $1.31$ PFLOPs, taking CAST from $8.72$ to $10.04$ PFLOPs per
target on CFA50. The trunk is not charged twice, since the pilot draws the same $K=4$
conditionings. Charged, CAST's margin over the $20\times36$ rung is $1.5\times$ rather than
$1.8\times$, and its $0.373$ still exceeds the best BoN allocation at or below $10.04$ PFLOPs,
$0.342$ at $8\times90$ (Table~\ref{tab:alloc}, Figure~\ref{fig:main4}(b)). CAST's advantage on hard
cases, in both performance and scaling, survives the charge.

\subsection{Dataset composition and filters}
\label{app:datasets}

CFA50 is screened from the CoFold Arena panel as described in \S\ref{sec:experiments}. The $80$
candidates entering that screen span $58$ Fab and $22$ VHH, $177$--$1024$ tokens (median $583$) and
$79$ antigens, and the $50$ retained span $38$ Fab and $12$ VHH, $177$--$880$ tokens (median $554$).

%Collected 21-hard is curated from PDB depositions instead, so it carries its own structural filters.
%A deposition is kept when it is an X-ray or cryo-EM structure below $3.5$\,\AA{} and its variable
%domains type under IMGT numbering with ANARCI. An antibody--antigen interface then counts when the
%two chains share a heavy-atom contact within $5$\,\AA{} over at least three antigen residues, and
%more than half the paratope falls inside the IMGT CDRs. The $21$ that survive the narrowing span
%$15$ Fab and $6$ VHH, $177$--$1155$ tokens (median $545$).

We curated two more dataset highlighting failure cases for the two cofolding model tested in our paper. From the PDB we filter for all antibody / antigen complexes released after 30 June 2025 that were novel relative to earlier entries. We retained X-ray and cryo-EM structures with resolution better than $3.5$,\AA{}, annotated variable domains using ANARCI under IMGT numbering, and required an antibody--antigen interface to satisfy three criteria: the chains had at least one heavy-atom contact within $5$,\AA{} involving at least three antigen residues; more than half of the paratope lay within the IMGT CDRs; and CDR-sequence clustering designated the complex as its cluster representative.

For these complexes, we performed BoN sampling for OpenDDE and Protenix V2. In both cases, we used 100 seeds to preselect the most difficult structures. The resulting 21-hard set was further filtered from the initial 100-seed Protenix V2 BoN run using ten times as many seeds, retaining only the 21 cases with DockQ scores below the CAPRI-medium threshold of 0.49. For the difficulty-balanced set for OpenDDE (\S\ref{app:opendde}), we selected 20 cases for which the initial 100-seed run produced no reasonable structure (e.g., DockQ $<0.23$), together with 20 cases for which the DockQ score exceeded 0.23.

\subsection{What a deeper unsteered pool still misses}
\label{app:failmode}

\begin{figure}[!ht]
\begin{center}
\includegraphics[width=\textwidth]{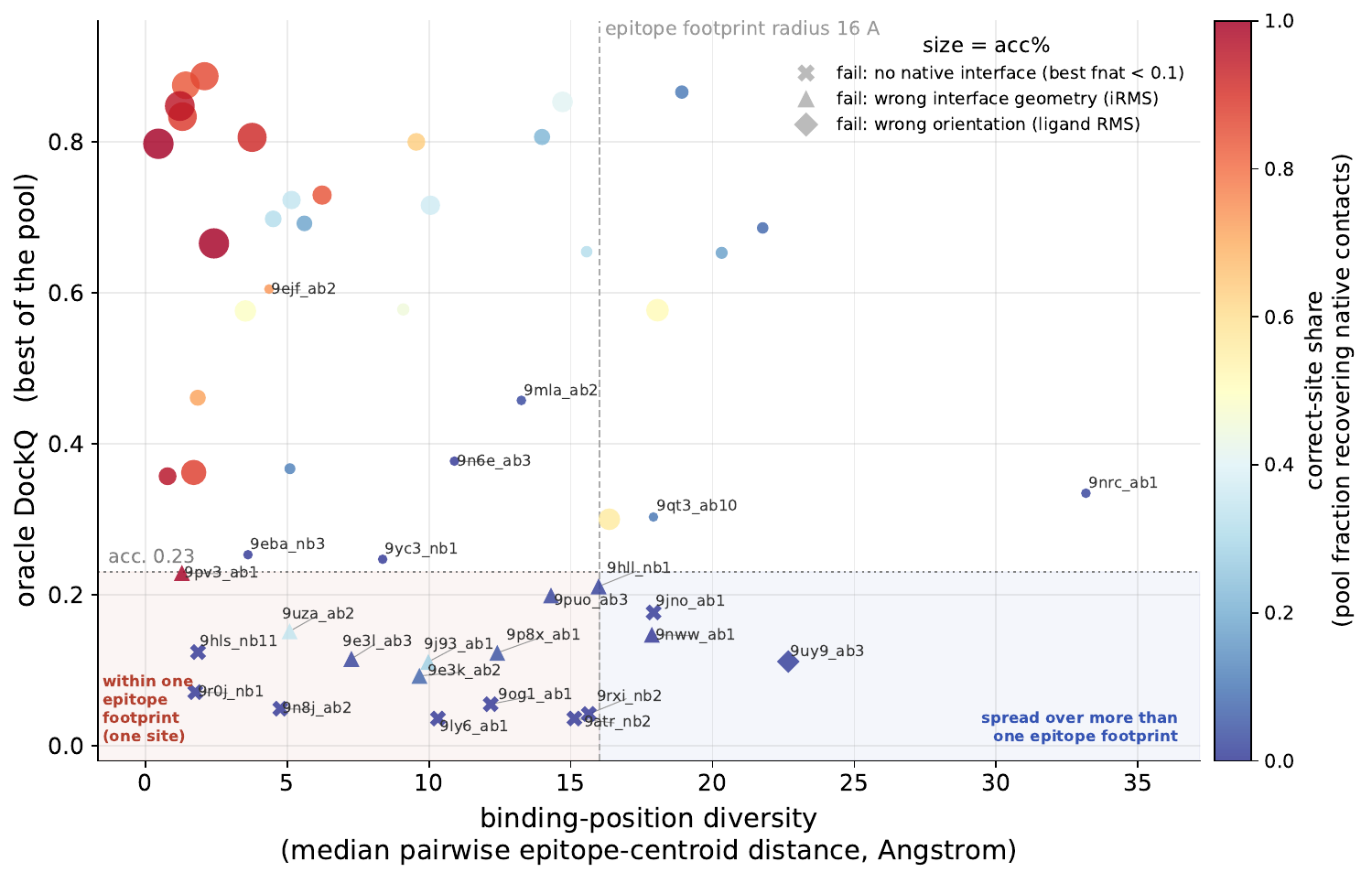}
\end{center}
\caption{Binding-position spread against the best $\mathrm{DockQ}$ in the pool, for the collected $52$
targets under the unsteered base model at $1{,}000$ structures each ($20$ subsamples $\times$ $50$
samples). Colour is the fraction of the pool recovering native contacts, marker area the fraction
reaching $\mathrm{DockQ}\ge0.23$, and marker shape what limits the best structure. The legend names
the three failure shapes, a cross for no native interface ($\mathrm{fnat}<0.1$), a triangle for
wrong interface geometry (iRMS) and a diamond for wrong orientation (ligand RMS), and a circle marks
a target with no such limit. Points below the dotted line hold nothing acceptable, so selection cannot help there.}
\label{fig:failmode}
\end{figure}

Figure~\ref{fig:failmode} characterises the failure this paper targets on a deeper pool. 
Every point is one target under the unsteered base model at $20$ alignment
subsamples $\times$ $50$ diffusion samples, so $1{,}000$ structures, $1.4\times$ the
$720$ used for the matched-budget comparisons.

The x-axis is how far apart the pool places the antibody, the median pairwise distance
between epitope centroids after each structure is put in a common antigen frame. The y-axis is
the best $\mathrm{DockQ}$ in the pool, i.e. $\mathrm{DockQ}^{\max}$. A target below the dotted line has no acceptable structure at
all, so no verifier applied to that pool can return one.

Three readings matter. First, $18$ of the pools sit below the line even at $1{,}000$ structures,
so the failure is not a budget that was too small. Second, the marker shape splits those $18$ by what
limits the best structure, and on $7$ it recovers no native contact at all, five of them with not one
draw in a thousand reaching the true epitope. That is a site failure, not a geometry one. Third, the
two sides of the dashed line separate two different site failures. Targets at the left, such as
9r0j\_nb1 and 9hls\_nb11 near $2$\,\AA{}, place the whole pool on one wrong patch, the redundant
binding mode of \citet{clifford2026pocket}. Targets at the right, such as 9rxi\_nb2 and 9jno\_ab1
above $15$\,\AA{}, spread across more than one epitope footprint and still miss. Spread alone
therefore does not fix a site, which is the case for a named destination, not more diversity.

\subsection{CFA50 target grouping}
\label{app:group}

Let $p$ be the fraction of $4$-conditioning draws whose $\mathrm{DockQ}^{\max}_{\mathrm{abag}}$
reaches $0.23$, at the same $4\times180$ allocation the comparison uses. A draw takes $4$ of the
target's conditionings and all $180$ samples of each, so $p$ is exact over all
$\binom{20}{4}=4845$ subsets. Then \emph{reliably fails} is $p\le0.10$, \emph{reliably solves} is $p\ge0.90$ and \emph{borderline}
the rest, giving $16$, $13$ and $21$ targets.

\subsection{The exact expectation over subsets}
\label{app:expect}

A rung that takes $m$ of the $n=180$ samples from each of $K$ conditionings could be drawn many
ways, and its $\mathrm{DockQ}^{\max}_{\mathrm{abag}}$ depends on which. We average over every draw
exactly, instead of picking one.

Fix a target and let $v_1<\dots<v_M$ be the distinct $\mathrm{DockQ}_{\mathrm{abag}}$ values in its
pool. Let $n_k(v)$ count the samples of conditioning $k$ at or below $v$. The maximum is at or below
$v$ only if every drawn sample is, and the conditionings are drawn independently, so
\begin{equation*}
\begin{split}
F(v)\;&=\;\Pr\big[\mathrm{DockQ}^{\max}_{\mathrm{abag}}\le v\big]
\;=\;\prod_{k=1}^{K}\binom{n_k(v)}{m}\Big/\binom{n}{m},\\[2pt]
\E\big[\mathrm{DockQ}^{\max}_{\mathrm{abag}}\big]\;&=\;\sum_{j=1}^{M}v_j\big(F(v_j)-F(v_{j-1})\big),
\end{split}
\end{equation*}
with $F(v_0)=0$ and $\binom{a}{b}=0$ for $a<b$. The sum is finite, so the rung carries no sampling
error. At $m=n$ the formula returns the pool maximum. The scaling curves of
Figure~\ref{fig:main4}(c,d) draw which conditionings to use instead, keeping every sample of each.
The same argument applies with $F(v)=\binom{C(v)}{c}\big/\binom{C}{c}$, where $c$ of the $C$
conditionings are drawn and $C(v)$ counts those whose own maximum is at or below $v$.

\subsection{BoN allocation sweep}
\label{app:alloc}

A matched budget can still be spent badly, and BoN in particular benefits more from seed scaling, so
we ask what the best use of $720$ unsteered structures is. Table~\ref{tab:alloc} holds the budget fixed and varies only the split. More seeds/conditionings is
monotonically better for BoN, worth $+0.044$ ($+13.7\%$) from $4$ to $20$. Table~\ref{tab:main} compares against the $4\times180$ arm because it matches
CAST's $K=4$ conditioning count. The higher-$K$ rungs score higher at more compute
(Figure~\ref{fig:main4}b). The rungs up to $20$ conditionings come from each target's own ladder at the matched $720$ structures.

\begin{table}[!ht]
\caption{Allocating a fixed $720$-structure budget for BoN, mean
$\mathrm{DockQ}^{\max}_{\mathrm{abag}}$ over the $50$ CFA50 targets. The structure budget is matched and the
compute budget is not, since the trunk runs once per conditioning, so PFLOPs rise with $K$ at a
constant $720$ structures. Both CAST and BoN use the same
trunk term, $1.82$ PFLOPs at $K=4$. Each rung uses a fixed conditioning set, the $K$ lowest indices. The $4\times180$ rung
takes every sample of its $4$ conditionings and is exact. Where a rung takes only part of a
conditioning we report the exact expectation of the maximum over all such subsets, a product of
hypergeometric terms per observed value (\S\ref{app:expect}). A fixed subset such as the first
$720/K$ samples would bias the rung, because sample index tracks confidence rank and the
confidence-best structures are the DockQ-worst.}
\label{tab:alloc}
\begin{center}
\small
\begin{tabular}{lrrr}
\toprule
\multicolumn{1}{c}{\bf ARM} &\multicolumn{1}{c}{\bf $K\times$ SAMPLES}
&\multicolumn{1}{c}{\bf PFLOPs/TGT} &\multicolumn{1}{c}{\bf MEAN $\mathrm{DockQ}^{\max}_{\mathrm{abag}}$} \\
\midrule
BoN      & $4\times180$  & $8.12$  & $0.322$ \\
BoN      & $8\times90$   & $9.94$  & $0.342$ \\
BoN      & $16\times45$  & $13.58$ & $0.351$ \\
BoN      & $20\times36$  & $15.40$ & ${\bf 0.366}$ \\
CAST             & $K=4$, $N=720$ & $8.72$ & ${\bf 0.373}$ \\
\bottomrule
\end{tabular}
\end{center}
\end{table}

\subsection{Candidate map reach}
\label{app:mapres}

\begin{table}[!ht]
\caption{Reach of the at most $S=15$ candidate map we ship, over the $50$ CFA50 targets, counting
targets where \emph{some} candidate centroid lands within $\delta$ of the true epitope centroid,
both taken on the same structure. Ground truth scores reach and never builds the map. The two
ingredient rows are the map's own modes and tiles, scored identically, so the hybrid row is their
combination and not a separate construction.}
\label{tab:map}
\begin{center}
\small
\begin{tabular}{lcccc}
\toprule
\multicolumn{1}{c}{\bf MAP} &\multicolumn{1}{c}{\bf $<6$ \AA}
&\multicolumn{1}{c}{\bf $<12$ \AA} &\multicolumn{1}{c}{\bf $<14$ \AA}
&\multicolumn{1}{c}{\bf MEDIAN (\AA)} \\
\midrule
Pool modes only                & $18$ & $31$ & $31$ & $9.39$ \\
Surface tiles only             & $25$ & $44$ & $48$ & $5.89$ \\
Hybrid, modes $+$ ranked tiles & ${\bf 32}$ & ${\bf 48}$ & ${\bf 49}$ & ${\bf 4.59}$ \\
\bottomrule
\end{tabular}
\end{center}
\end{table}

The shipped map is the hybrid, pool modes first and then surface tiles at a $12$-tile budget, ranked
by their overlap with those modes. Its closest candidate is a median $4.59$\,\AA{} from the true
epitope (Table~\ref{tab:map}). Modes and tiles fail on different targets, which is why the hybrid
uses both. Scored apart, the tiles carry the coverage and the modes the centring. Tiles alone reach
$25$ targets within $6$\,\AA{} against the modes' $18$, and adding the modes takes the hybrid to
$32$ and the median from $5.89$ to $4.59$\,\AA{}. The hybrid is closer than tiles alone on $26$ of
the $50$ targets and further on $3$.

\begin{figure}[!ht]
\begin{center}
\includegraphics[width=0.52\textwidth]{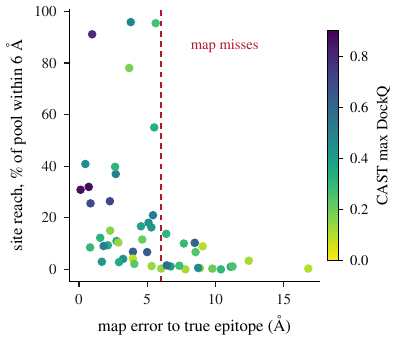}
\end{center}
\caption{Site reach, the share of the pool whose predicted epitope centroid lands within
$6$~\AA{} of the true one, against map error, the distance from the best map candidate to the true
epitope. Colour is CAST $\mathrm{DockQ}^{\max}_{\mathrm{abag}}$, darker is higher. The dashed line at
$6$~\AA{} is where the map is counted as missing the epitope.}
\label{fig:map}
\end{figure}

Figure~\ref{fig:map} separates a map failure from a method failure. Every one of the $32$ targets
whose map lands within $6$\,\AA{} achieves non-zero site reach, so a good map always puts some of
the pool on the true site. Reach reports only that the site was visited. The pose found there must
also be acceptable, and $7$ of those $32$ fall short of $0.23$. The targets to the right of the line
fail for the other reason, that the map itself misses.

The map does not cap where CAST arrives. On one target CAST produced a structure $4.3$\,\AA{} from
the true epitope through a map whose best candidate was $12.3$\,\AA{} out, while BoN never came
within $10$\,\AA{} in $720$ draws. The tilt supplies a direction, and diffusion supplies the rest.

\subsection{Ablations}
\label{app:abl}

Each ablation changes one field of the reference arm, holding $K=4$ and $N=720$. \textbf{A1}
replaces the joint pool with a separate model per conditioning, the arrangement TDS
\citep{wu2023practical} uses. Each conditioning gets its own $720$-particle pool, so A1 draws
$4\times720$ structures where CAST draws $720$. Splitting one $720$ budget four ways would leave
three particles per cell instead of twelve, so the separate arrangement pays per conditioning to
keep the site sweep. \textbf{A2} removes the gradient and \textbf{A3} collapses the umbrella sweep
to $d^\star=12$\,\AA{}. \textbf{A4} substitutes the oracle map, the true epitope read from the
ground-truth complex. That leaves one candidate per target in place of the fifteen, so every
particle is aimed at the right site. \textbf{A5} substitutes a site-shuffled map and \textbf{A6}
disables resampling. All run on all $50$ CFA50 targets.

\begin{table}[!ht]
\caption{Ablations at $K=4$, $N=720$ over all CFA50 targets, split by the groups of
Table~\ref{tab:main}. A1 is the mean over its four conditionings with the per-target standard
deviation beneath, a spread CAST avoids by pooling all four into one run.}
\label{tab:abl}
\begin{center}
\small
\setlength{\tabcolsep}{8pt}
\begin{tabular}{llrrrrr}
\toprule
\multicolumn{1}{c}{\bf ARM} &\multicolumn{1}{c}{\bf CHANGE}
&\multicolumn{4}{c}{\bf MEAN $\mathrm{DockQ}^{\max}_{\mathrm{abag}}$}
&\multicolumn{1}{c}{\bf REACH $0.23$} \\
 & &\multicolumn{1}{c}{all ($50$)} &\multicolumn{1}{c}{fails ($16$)}
&\multicolumn{1}{c}{border ($13$)} &\multicolumn{1}{c}{solves ($21$)} & \\
\midrule
CAST & reference       & $0.373$ & $0.182$ & $0.385$ & $0.511$ & $36$ \\
\addlinespace
A1 & separate pools    & $0.367$ & $0.193$ & $0.397$ & $0.481$ & $33.2$ \\
 &   \;\;\emph{sd}     & $\pm0.075$ & $\pm0.057$ & $\pm0.107$ & $\pm0.068$ & \\
A2 & no gradient       & $0.310$ & $0.127$ & $0.200$ & $0.518$ & $28$ \\
A3 & one radius        & $0.361$ & $0.178$ & $0.378$ & $0.490$ & $35$ \\
A4 & oracle map        & ${\bf 0.484}$ & ${\bf 0.345}$ & ${\bf 0.447}$ & ${\bf 0.614}$ & ${\bf 42}$ \\
A5 & shuffled map      & $0.319$ & $0.167$ & $0.249$ & $0.477$ & $28$ \\
A6 & no resampling     & $0.373$ & $0.221$ & $0.351$ & $0.502$ & $38$ \\
\bottomrule
\end{tabular}
\end{center}
\end{table}

\textbf{The joint pool buys robustness and cost.}
\label{sec:res-tds}
A1 matches CAST on every group, so the level comes from the tilt, which both arms share. What the joint pool supplies is certainty and cost. A separate run commits to one conditioning, and
its spread across conditionings is $\pm0.075$, where pooling them moves the mean by only $0.006$. The
draw dominates by an order of magnitude. On $33$ of $50$ targets it decides whether such a run
matches CAST. On $15$ of $50$ it decides whether the run returns an acceptable structure at all. CAST
returns the pooled result every time, from one pool where the separate arrangement needs $K$, at $8.72$ PFLOPs against about
$29$ (\S\ref{sec:cast}).

\textbf{The gradient and the map act where the site is uncertain.} The gradient is worth $+0.185$
($+92.5\%$) on borderline and $+0.055$ ($+43.3\%$) on fails, the candidate map against shuffled
(A5) $+0.136$ ($+54.6\%$) and
$+0.015$ ($+9.0\%$). Each carries $8$ of the pooled reach gain, $6$ of it on borderline. The gradient
also supplies $3$ of the $4$ fails targets CAST reaches at all, where the candidate map supplies $1$. On
the targets BoN already solves both are small, $-0.007$ ($-1.4\%$) and $+0.034$ ($+7.1\%$).

\textbf{Resampling trades the fails group for the borderline one.} Resampling concentrates the pool
where the reward already scores well. It gains $+0.034$ ($+9.6\%$) on borderline and $+0.009$
($+1.7\%$) on solves. It pays $0.039$ ($-17.5\%$) on fails, reaching $4$ of $16$ there against $7$
without it. The three cancel in the pooled mean, which a paired test over the $50$ cannot separate
from zero ($p=1.00$). FK, which resamples without the gradient, makes the same trade
(Table~\ref{tab:baselines}). CAST therefore resamples early and only twice,
with a single tilted event. That takes the borderline gain while leaving the fails group a pool that
is still exploring.

\textbf{The umbrella sweep helps consistently and by a small amount.} $4$ radii beat one in all three group
means and in reach, $36$ targets against $35$, and on $29$ of the $50$ targets, the largest gain being
$+0.021$ ($+4.2\%$) where BoN already solves. No group is harmed. The pooled gain of $+0.012$
($+3.3\%$) is small enough to trade away when the budget is tight, since one umbrella holds the same
$12$ particles per cell at $N=180$ that four need $N=720$ to reach. We keep the sweep on the
consistency of its sign.

\textbf{The map sets the ceiling without capping it.} A4 substitutes the oracle map and adds $+0.163$
($+89.6\%$) on fails, taking reach from $36$ to $42$, more than any change to the search. CAST's
ceiling is set by where it is told to look. The map is not a bound, though. The shipped one reaches
$32$ of the $50$ targets within $6$\,\AA{} and $48$ within $12$\,\AA{}, yet CAST solves targets whose
best candidate is further out, so the tilt biases the search while leaving it free to depart
(\S\ref{app:mapres}).

\subsection{Transfer to a second co-folding model}
\label{app:opendde}

CAST is backbone-agnostic by construction, so the question is whether the gain survives a change of
base model. We repeat the matched-budget comparison on \textbf{OpenDDE} \citep{opendde2026}, an independently trained
all-atom co-folding model with an antibody-antigen fine-tuned checkpoint, over $40$ difficulty-balanced targets from the public PDB (\S\ref{app:datasets}). \textit{No
hyperparameter is retuned}. The guidance scale and cap, the umbrella sweep, the resampling scheme
and window, the level correction and the $720$ structures as $4\times180$ are the values CFA50
uses, and the map is built by the same reference-free construction. What changes is the base
model, which is the experiment, so its weights are its own and the map is a different instance
over a different target set. Targets are grouped
by BoN's own reliability exactly as in \S\ref{app:group}, $p$ exact over all $\binom{20}{4}=4845$
four-conditioning draws, giving $16$ reliably-failing, $7$ borderline and $17$ reliably-solving
targets.

At the matched budget CAST reaches $0.454$ against BoN's $0.372$ (Table~\ref{tab:opendde-main}), a difference of $+0.082$, and
brings $27$ of the $40$ targets to an acceptable pose against $20$. The gain sits where it does on
CFA50. On the $16$ targets BoN reliably fails it is $+0.111$, improving $16$ of $16$ and taking $6$
targets across the threshold from none. Where BoN already solves, the difference is $+0.011$ and a
paired test cannot separate it from zero. The borderline group is $7$ targets, too few to carry a
test of its own, and its mean rises $+0.187$ where its median rises $+0.087$, with two targets supplying most of that.

\begin{table}[!ht]
\caption{$\mathrm{DockQ}^{\max}_{\mathrm{abag}}$ over the $40$ OpenDDE40 targets for CAST and BoN,
both at a \emph{matched} budget of $720$ structures and the same $4$ conditionings, BoN spending it
as $4\times180$. The counterpart of Table~\ref{tab:main}, with the grouping built the same way
(\S\ref{app:group}) from BoN alone. SITE is as in Table~\ref{tab:main}. REACH $0.23$ counts targets attaining
$\mathrm{DockQ}^{\max}_{\mathrm{abag}}\ge0.23$, CAST WINS is a paired count spanning both rows, and
$p$ is a two-sided Wilcoxon signed-rank test on the per-target differences within that subset.}
\label{tab:opendde-main}
\begin{center}
\small
\setlength{\tabcolsep}{3.2pt}
\begin{tabular}{llrrrrr}
\toprule
\multicolumn{1}{c}{\bf SUBSET} &\multicolumn{1}{c}{\bf ARM}
&\multicolumn{1}{c}{\bf MEAN $\mathrm{DockQ}^{\max}_{\mathrm{abag}}$}
&\multicolumn{1}{c}{\bf SITE (\AA{})}
&\multicolumn{1}{c}{\bf REACH $0.23$}
&\multicolumn{1}{c}{\bf CAST WINS}
&\multicolumn{1}{c}{\bf $p$} \\
\midrule
\multirow{2}{*}{All targets ($40$)}
  & CAST & ${\bf 0.454}$ & ${\bf 1.75}$ & ${\bf 27}$ & \multirow{2}{*}{$29$} & \multirow{2}{*}{$<10^{-4}$} \\
  & BoN  & $0.372$ & $6.41$ & $20$ & & \\
\addlinespace
\multirow{2}{*}{BoN reliably fails ($16$)}
  & CAST & ${\bf 0.211}$ & ${\bf 3.07}$ & ${\bf 6}$ & \multirow{2}{*}{$16$} & \multirow{2}{*}{$<10^{-4}$} \\
  & BoN  & $0.101$ & $13.88$ & $0$ & & \\
\addlinespace
\multirow{2}{*}{Borderline ($7$)}
  & CAST & ${\bf 0.372}$ & ${\bf 1.75}$ & ${\bf 4}$ & \multirow{2}{*}{$6$} & \multirow{2}{*}{$0.11$} \\
  & BoN  & $0.186$ & $2.87$ & $3$ & & \\
\addlinespace
\multirow{2}{*}{BoN reliably solves ($17$)}
  & CAST & $0.715$ & ${\bf 0.51}$ & $17$ & \multirow{2}{*}{$7$} & \multirow{2}{*}{$0.98$} \\
  & BoN  & $0.704$ & $0.85$ & $17$ & & \\
\bottomrule
\end{tabular}
\end{center}
\end{table}

\begin{figure}[!ht]
\begin{center}
\includegraphics[width=\textwidth]{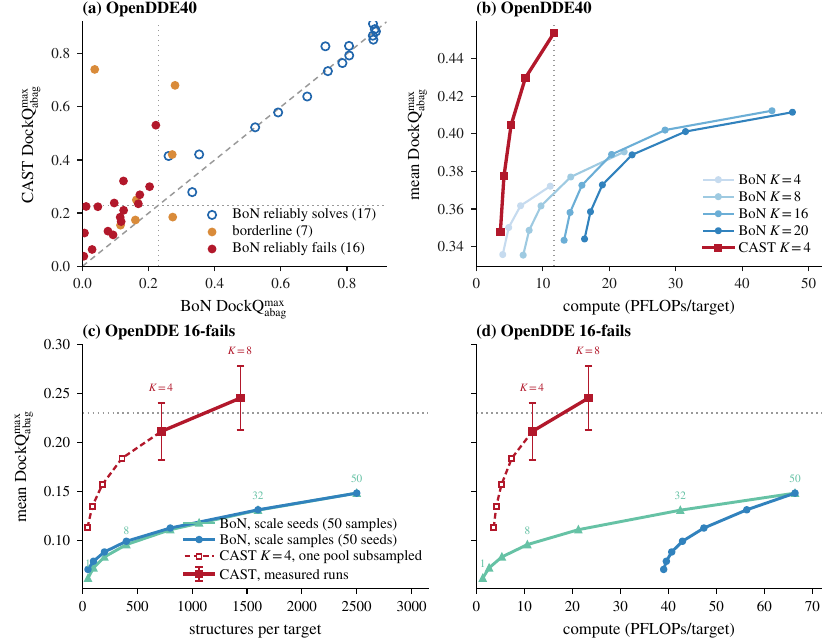}
\end{center}
\caption{OpenDDE40, the counterpart of Figure~\ref{fig:main4}. \textbf{(a)} Per-target
$\mathrm{DockQ}^{\max}_{\mathrm{abag}}$, both arms at $720$ structures and the same $4$ conditionings,
fill encoding the grouping. \textbf{(b)} The same set against compute, BoN drawn at each
conditioning count. \textbf{(c)} and \textbf{(d)} are the scaling ladder on the $16$ targets BoN
reliably fails, against structures and compute, drawn as in Figure~\ref{fig:main4}(c,d).
Best-of-$N$ appears twice, scaling conditionings at $50$ samples each and scaling samples at $50$
conditionings, each the exact expectation over subsets of that size. Filled squares are measured
CAST runs at $K=4$ and $K=8$ along $N=180K$, with bars giving the standard error over the $16$
targets. Open squares on the dashed line are the $K=4$ pool of $720$ drawn at fewer particles.}
\label{fig:opendde4}
\end{figure}

The compute axis separates the two backbones. On CFA50 CAST matches BoN's best allocation of the
budget, whereas here it exceeds every allocation at $11.68$ PFLOPs per target, and BoN does not 
close the gap by spreading the same budget over
more conditionings. Even at $20\times36$, $5\times$ the conditionings and $2\times$ the compute at
$23.43$ PFLOPs per target, BoN reaches $0.389$ against CAST's $0.454$
(Table~\ref{tab:opendde-alloc}, Figure~\ref{fig:opendde4}b).

\begin{table}[!ht]
\caption{Allocating a fixed $720$-structure budget on OpenDDE40, mean
$\mathrm{DockQ}^{\max}_{\mathrm{abag}}$ over the $40$ targets, the counterpart of
Table~\ref{tab:alloc}. Conditionings are the $K$ lowest indices and partial rungs are the exact
expectation over sample subsets (\S\ref{app:expect}).}
\label{tab:opendde-alloc}
\begin{center}
\small
\begin{tabular}{lrrr}
\toprule
\multicolumn{1}{c}{\bf ARM} &\multicolumn{1}{c}{\bf $K\times$ SAMPLES}
&\multicolumn{1}{c}{\bf PFLOPs/TGT} &\multicolumn{1}{c}{\bf MEAN $\mathrm{DockQ}^{\max}_{\mathrm{abag}}$} \\
\midrule
BoN      & $4\times180$  & $11.12$ & $0.372$ \\
BoN      & $8\times90$   & $14.20$ & $0.377$ \\
BoN      & $16\times45$  & $20.36$ & $0.389$ \\
BoN      & $20\times36$  & $23.43$ & $0.389$ \\
CAST     & $K=4$, $N=720$ & $11.68$ & ${\bf 0.454}$ \\
\bottomrule
\end{tabular}
\end{center}
\end{table}

On the $16$ targets BoN reliably fails, CAST's smallest rung already beats the largest Best-of-$N$
pool we ran (Figure~\ref{fig:opendde4}(c,d)). CAST reaches $0.211$ at $720$ structures and crosses the $0.23$ acceptable line
at $1440$ with $0.245$, a step of $+0.034$ at its standard error, where Best-of-$N$ climbs
logarithmically along either axis to only $0.148$ at $2500$ structures and never reaches it. Compute widens the gap, since
Best-of-$N$ pays a trunk pass per conditioning and needs $66.4$ PFLOPs for that $0.148$ while CAST
spends $11.7$ for $0.211$. No Best-of-$N$ allocation on either axis reaches CAST at any budget in
the figure. This reproduces the pattern Protenix shows in Figure~\ref{fig:main4}(c,d), where CAST's
$K=4$ rung also passes Best-of-$N$'s largest pool at a fifth of the compute.

\end{document}

%% file: math_commands.tex
\usepackage{amsmath,amsfonts,bm}

\def\eqref#1{equation~\ref{#1}}
\def\1{\bm{1}}

\DeclareMathAlphabet{\mathsfit}{\encodingdefault}{\sfdefault}{m}{sl}
\SetMathAlphabet{\mathsfit}{bold}{\encodingdefault}{\sfdefault}{bx}{n}

\newcommand{\E}{\mathbb{E}}

%% file: commands.tex
\renewcommand{\E}{\mathbb{E}} % expectation
\newcommand{\g}{\,|\,} % given
\newcommand{\norm}[1]{\left\lVert#1\right\rVert}
\newcommand{\reals}{\mathbb{R}} % real line
\newcommand{\I}{\mathbb{I}}

\newcommand{\normal}{\mathcal{N}}